\documentclass[twocolumn,trackchanges,resetfootnote]{aastex7}
\usepackage{CJK} 

\renewcommand{\thefootnote}{\dag\arabic{footnote}}

\shortauthors{Kakimoto et al.}
\DeclareUnicodeCharacter{2061}{}
\begin{document}
\begin{CJK*}{UTF8}{bsmi} 

\title{DeepDive: Simultaneous Formation of Massive Quiescent Galaxies in High-Redshift Galaxy Overdensities}

\author[0000-0003-2918-9890]{Takumi Kakimoto}
\affiliation{Department of Astronomical Science, The Graduate University for Advanced Studies, SOKENDAI, 2-21-1 Osawa, Mitaka, Tokyo 181-8588, Japan}
\affiliation{National Astronomical Observatory of Japan, 2-21-1 Osawa, Mitaka, Tokyo 181-8588, Japan}
\email[show]{takumi.kakimoto@grad.nao.ac.jp}  

\author[0000-0002-5011-5178]{Masayuki Tanaka}
\affiliation{National Astronomical Observatory of Japan, 2-21-1 Osawa, Mitaka, Tokyo 181-8588, Japan}
\affiliation{Department of Astronomical Science, The Graduate University for Advanced Studies, SOKENDAI, 2-21-1 Osawa, Mitaka, Tokyo 181-8588, Japan}
\email{masayuki.tanaka@nao.ac.jp} 

\author[0000-0002-9453-0381]{Kei Ito}
\affiliation{Cosmic Dawn Center (DAWN), Denmark}
\affiliation{DTU Space, Technical University of Denmark, Elektrovej 327, 2800 Kgs. Lyngby, Denmark}
\email{keiit@dtu.dk} 

\author[0000-0001-6477-4011]{Francesco Valentino}
\affiliation{Cosmic Dawn Center (DAWN), Denmark}
\affiliation{DTU Space, Technical University of Denmark, Elektrovej 327, 2800 Kgs. Lyngby, Denmark}
\email{fmava@dtu.dk} 

\author[0000-0002-4225-4477]{Makoto Ando}
\affiliation{Institute for Cosmic Ray Research, The University of Tokyo, 5-1-5 Kashiwanoha, Kashiwa, Chiba 277-8582, Japan}
\affiliation{National Astronomical Observatory of Japan, 2-21-1 Osawa, Mitaka, Tokyo 181-8588, Japan}
\email{makoto.ando.astro@gmail.com} 

\author[0000-0003-2680-005X]{Gabriel Brammer}
\affiliation{Cosmic Dawn Center (DAWN), Denmark}
\affiliation{Niels Bohr Institute, University of Copenhagen, Jagtvej 128, DK2200 Copenhagen N, Denmark}
\email{gabriel.brammer@nbi.ku.dk}

\author[0000-0001-6763-5551]{Massissilia L. Hamadouche}
\affiliation{Department of Astronomy, University of Massachusetts, Amherst, MA 01003, USA}
\email{mhamadouche@umass.edu}

\author[0000-0002-5588-9156]{Vasily Kokorev}
\affiliation{Department of Astronomy, The University of Texas at Austin, Austin, TX 78712, USA}
\email{vkokorev@utexas.edu}

\author[0000-0002-0243-6575]{Jacqueline Antwi-Danso}\thanks{Dunlap Fellow}
\affiliation{David A. Dunlap Department of Astronomy \& Astrophysics, University of Toronto, 50 St George Street, Toronto, ON M5S 3H4, Canada}
\affiliation{Dunlap Institute for Astronomy and Astrophysics, 50 St. George Street, Toronto, Ontario, M5S 3H4, Canada}
\email{j.antwidanso@utoronto.ca}

\author[0000-0003-0215-1104]{William M. Baker}
\affiliation{DARK, Niels Bohr Institute, University of Copenhagen, Jagtvej 155A, DK-2200 Copenhagen, Denmark}
\email{william.baker@nbi.ku.dk}

\author[0000-0002-8680-248X]{Daniel Ceverino}
\affiliation{Departamento de Fisica Teorica, Modulo 8, Facultad de Ciencias, Universidad Autonoma de Madrid, 28049 Madrid, Spain}
\affiliation{CIAFF, Facultad de Ciencias, Universidad Autonoma de Madrid, 28049 Madrid, Spain}
\email{daniel.ceverino@uam.es}

\author[0000-0002-9382-9832]{Andreas L. Faisst}
\email{afaisst@caltech.edu}
\affiliation{IPAC, California Institute of Technology, 1200 E. California Blvd. Pasadena, CA 91125, USA}

\author[0000-0002-5228-2244]{Marion Farcy}
\affiliation{Institute for Physics, Laboratory for Galaxy Evolution and Spectral modelling, Ecole Polytechnique Federale de Lausanne, Observatoire de Sauverny, Chemin Pegasi 51, 1290 Versoix, Switzerland}
\email{marion.farcy@epfl.ch}

\author[0000-0002-3301-3321]{Michaela Hirschmann}
\affiliation{Institute for Physics, Laboratory for Galaxy Evolution and Spectral modelling, Ecole Polytechnique Federale de Lausanne, Observatoire de Sauverny, Chemin Pegasi 51, 1290 Versoix, Switzerland}
\email{michaela.hirschmann@epfl.ch}

\author[0000-0002-8896-6496]{Christian Kragh Jespersen}
\affiliation{Department of Astrophysical Sciences, Princeton University, Princeton, NJ 08544, USA}
\email{ckragh@princeton.edu}

\author[0000-0002-7598-5292]{Mariko Kubo}
\affiliation{Department of Physics and Astronomy, School of Science, Kwansei Gakuin University, 1 Gakuen Uegahara, Sanda, Hyogo 669-1330, Japan}
\affiliation{Astronomical Institute, Tohoku University, 6-3, Aramaki, Aoba, Sendai, Miyagi 980-8578, Japan}
\email{markubo@kwansei.ac.jp} 

\author[0000-0003-2475-124X]{Allison W. S. Man}
\affiliation{Department of Physics \& Astronomy, University of British Columbia, 6224 Agricultural Road, Vancouver, BC V6T 1Z1, Canada}
\email{aman@phas.ubc.ca}

\author[0000-0003-3228-7264]{Masato Onodera}
\affiliation{Subaru Telescope, National Astronomical Observatory of Japan, National Institutes of Natural Sciences (NINS), 650 North A'ohoku Place, Hilo, HI 96720, USA}
\affiliation{Department of Astronomical Science, The Graduate University for Advanced Studies, SOKENDAI, 2-21-1 Osawa, Mitaka, Tokyo 181-8588, Japan}
\email{monodera@naoj.org} 

\author[0000-0003-4442-2750]{Rhythm Shimakawa}
\affiliation{Waseda Institute for Advanced Study (WIAS), Waseda University, 1-21-1, Nishi-Waseda, Shinjuku, Tokyo 169-0051, Japan}
\email{rhythm.shimakawa@aoni.waseda.jp}

\author[0000-0003-3631-7176]{Sune Toft}
\affiliation{Cosmic Dawn Center (DAWN), Denmark}
\affiliation{Niels Bohr Institute, University of Copenhagen, Jagtvej 128, DK2200 Copenhagen N, Denmark}
\email{sune@nbi.ku.dk} 

\author[0000-0003-1614-196X]{John R. Weaver}\thanks{Brinson Prize Fellow}
\affiliation{MIT Kavli Institute for Astrophysics and Space Research, 70 Vassar Street, Cambridge, MA 02139, USA}
\email{john.weaver.astro@gmail.com}

\author[0000-0002-9665-0440]{Po-Feng Wu}
\affiliation{Graduate Institute of Astrophysics, National Taiwan University, Taipei 10617, Taiwan}
\affiliation{Department of Physics and Center for Theoretical Physics, National Taiwan University, Taipei 10617, Taiwan}
\email{wupofeng@phys.ntu.edu.tw}

\author[0000-0002-6768-8335]{Pengpei Zhu (朱芃佩)}
\affiliation{Cosmic Dawn Center (DAWN), Denmark}
\affiliation{DTU Space, Technical University of Denmark, Elektrovej 327, 2800 Kgs. Lyngby, Denmark}
\email{penzhu@space.dtu.dk}


\begin{abstract}

We report on the spectroscopic confirmation of overdense regions of massive quiescent galaxies (QGs) in the early Universe with JWST/NIRSpec. Based on data from the \textit{DeepDive} NIRSpec program and archival data from the Dawn JWST Archive, we confirm three QGs in the vicinity of \textit{Jekyll \& Hyde}, a pair of massive QG and a dusty star-forming galaxy, at $z=3.71$ and two QGs around SXDS-27434 at $z=4.01$. According to the analysis of galaxy number density with photometric redshifts, \textit{Jekyll \& Hyde} (SXDS-27434) are in an overdense region, where the number density of galaxies is three (four) times higher than the average in the COSMOS (SXDS) field. SED fitting suggests that most of the QGs follow similar star formation histories and have consistent formation and quenching epochs. The same trend is observed in other proto-clusters hosting QGs that were already identified by ground-based telescopes, indicating that the large-scale environment plays an important role in the formation of QGs. In addition, JWST spectra reveal a broad H$\alpha$ emission line from SXDS-27434 and faint emission lines from other three QGs, which are identified as AGN-driven based on their emission line ratios. The overdensity is also reproduced by the Illustris TNG300 simulation at $z=3.71$, in which the member QGs also have similar quenching epochs. These results are consistent with a scenario in which the large-scale structure enhances merger activity and/or gas accretion and triggers AGN feedback, thereby driving simultaneous formation and quenching of QGs in the overdensity.

\end{abstract}



\section{Introduction} 
The distribution of galaxies in the Universe exhibits significant fluctuations, forming the ``large-scale structure'' \citep[e.g.,][]{Geller1989,Doroshkevich2004, Gott2005}.
The large-scale structure has now been detected out to $z\sim10$ through both clustering \citep{Paquereau.etal.2025} and field-to-field fluctuations \citep[known as cosmic variance,][]{WeibelAndJespersen.etal.2025}.
The physical properties of galaxies depend on their surrounding environments. For instance, elliptical galaxies dominate clusters of galaxies in the local Universe \citep[e.g.,][]{Dressler1980}. They tend to have red colors, especially in the cluster cores \citep[forming the ``red sequence'' on a color-magnitude diagram; e.g.,][]{Bower1992,Hogg2004}. While the tile of the red sequence is primally driven by metallicity, its small color scatter indicates a narrow age differences among the member galaxies \citep[e.g.,][]{Bower1992,Kodama1998,Bower1998}.
These clusters are also confirmed in the distant Universe (at $z<2$) and show the red sequence, suggesting that there are evolved galaxies in distant clusters \citep[e.g.,][]{Muzzin2009,Wilson2009, Tanaka2010, Willis2020}.

To gain deeper insights into the formation and evolution of galaxies in clusters, previous studies confirm distant proto-clusters, which are overdense regions of galaxies but are not necessarily gravitationally bound systems. They are likely progenitors of local clusters \citep[][for a review]{Overzier_2016}.
Unlike the virialized clusters in the local Universe, it is difficult to detect them via diffuse X-ray emission and/or the Sunyaev-Zel'dovich effect \citep[e.g.,][]{Sunyaev1972,Blakeslee2003,Brodwin2010,Gozaliasl2019}. Thus, previous studies have widely searched for overdense regions using galaxy distributions, which trace the underlying matter distribution in the Universe. In addition, constraining their redshift information (i.e., line-of-sight positions) requires homogeneous, wide-field, and complete observations, but few instruments exist that can achieve complete spectroscopic follow-up for all galaxies. Thanks to recent wide-field photometric surveys, such as the Cosmic Evolution Survey \citep[COSMOS;][]{Scoville2007}, we are now able to define the environment while ensuring mass completeness based on accurate photometric redshifts. This approach allows us to trace the underlying, unbiased matter density distribution over a wide area, which cannot be fully captured by biased tracers like $\rm Ly\alpha$ emitters \citep[e.g.,][]{Shimakawa2017,Ito2021,Yonekura2022}.

Unlike clusters in the local Universe, proto-clusters often contain (dusty) star-forming galaxies (SFGs) and starburst galaxies \citep[e.g.,][]{Miller2018,Oteo2018,Mitsuhashi2021,Lemaux2022,Sun2026}. Thanks to the recent deployment of the James Webb Space Telescope (JWST), these proto-clusters have now been confirmed up to $z\sim 8$ \citep[e.g.,][]{Morishita2023, Helton2024}. These findings indicate that galaxy proto-clusters see reverse star formation activity compared with clusters at low redshift and suggest increased star formation rate in dense environment at high redshift ($z>2$) due to cold accretion in massive halos \citep[e.g.,][]{Keres2005, Dekel2006}.
Interestingly, proto-clusters with several quiescent galaxies (QGs) have also been found up to $z\sim 4$ \citep{McConachie2022, Ito2023, Tanaka2024, Jespersen2025}, demonstrating the diversity of star formation activities. Although there seems to be more abundant gas in dense environments compared with the field, these findings suggest the lower star formation activity in some proto-clusters. Thus, how the surrounding environment affects star formation in the distant Universe still remains a key open question.

Massive elliptical galaxies in the local Universe are thought to be formed in the early Universe \citep[at $3\lesssim z \lesssim 5$, e.g.,][]{Nelan_2005,Thomas_2005,Thomas2010,Renzini_2006}, and direct observations of their physical properties close to their formation epochs are crucial to understand the formation process. In fact, recent near-infrared spectroscopic observations confirm such recently formed QGs at high redshifts \citep[e.g.,][]{Schreiber_2018,Forrest_2020,Valentino2020,Setton2024,Wu2025,Nanayakkara2025,DeepDive_2025,Baker2025_FLAMINGO,Baker2025_MF,Hamadouche2026}.

With the advent of JWST, the confirmation of broad Balmer lines reveal the presence of active galactic nuclei (AGN), and Na \textsc{i} Doublet absorption features (tracing outflows) from the spectra of massive QGs suggests that AGN feedback plays a key role in quenching \citep[e.g.,][]{carnall_2023, Eugenio2024, Belli2024, Davies2024, Valentino_2025, Zhu2026}. In addition, studies investigating the surrounding environment of these QGs argue for the importance of their large halo masses and merger activity, as they are located in overdense regions \citep[e.g.,][]{Kakimoto2024, Carnall2024, Jespersen2025, deGraaff2025, Ito2025b, McConachie2025,Baker2026}. However, the discovery of these overdensities has largely relied on star-forming galaxy (candidates), making them strongly biased toward line emitters and Lyman-break galaxies. Therefore, to fully understand the variations in star formation activity within proto-clusters at high redshifts, it is essential to identify proto-clusters using all galaxy populations, including QGs. A recent JWST Cycle 2 GO program, \textit{DeepDive} \citep[PID \#3567, PI: F. Valentino;][]{DeepDive_2025}, has spectroscopically confirmed 24 massive QGs at $3<z<5$ using medium resolution spectra, enabling us to statistically investigate the evolutionary processes of QGs. Utilizing these spectra provides a highly efficient approach to tightly constrain their star formation histories (SFHs).

In this paper, we report the confirmation of two overdensities of massive QGs at $z=3.71$ and $4.01$ in the COSMOS and SXDS fields with JWST/NIRSpec. The QGs exhibit similar spectral shapes, suggesting similar SFHs. We thus investigate the similarity of the SFHs of the QGs in overdensities with spectral modeling and compare it to that of the field QGs to characterize the role of environment in the formation and quenching of these massive systems. 

This paper is structured as follows. First, we introduce the targets and environmental density estimations using wide-field survey data in Section \ref{sec:env}. Then we introduce the spectroscopic follow-up observations made with JWST, as well as the photometric measurements used for constraining the physical properties of individual galaxies in Section \ref{sec:obs}. In Section \ref{sec:Results}, the physical properties of the QGs inferred from their spectral energy distribution (SED) fitting are summarized. In Section \ref{sec:Discussion}, we discuss the unique properties of overdensity members as well as the implications of our findings. Finally, we conclude the paper in Section \ref{sec:conclusion}. We assume the \cite{Chabrier_2003} initial mass function (IMF) and a flat $\mathrm{\Lambda CDM}$ cosmology with $H_0 = 70\,\mathrm{km\,s^{-1}\,Mpc^{-1}}, \Omega_m = 0.3$, and $\Omega_\Lambda = 0.7$. All magnitudes are in the AB system \citep{Oke1983}. 

\section{Targets and Density Estimates} \label{sec:env}

\subsection{Targets} \label{sec:target}
In this study, we examine the environments of two massive QGs at high redshift. The first target is the first massive QG confirmed at $z>3.5$ and its dusty star-forming companion, the ``\textit{Jekyll \& Hyde}'' pair \citep[$z=3.71$;][]{Glazebrook2017, Simpson2017, Jekyll2018, Schreiber_2018}. The massive QG, \textit{Jekyll} ($M_*\sim 10^{11}M_\odot,\mathrm{SFR}_{\mathrm{SED}}<0.2\, M_\odot\, \mathrm{yr^{-1}}$), was initially confirmed with Keck/MOSFIRE spectroscopy \citep{McLean2010,McLean2012} and subsequently targeted by extensive follow-up campaigns, including NIRCam imaging from the PRIMER \citep[PID \#1837, PI: J. Dunlop;][]{Donnan_2024} and COSMOS-Web surveys \citep[PID \#1727, PI: J. Kartaltepe;][]{Casey_2023}, NIRSpec/MSA PRISM spectroscopy \citep[PID \#2565, PI: K. Glazebrook;][]{Nanayakkara2025}, G235M/F170LP medium-resolution slit spectroscopy \citep[\textit{DeepDive};][]{DeepDive_2025}, and G235H high-resolution IFU observations \citep[PID \#1217, PI: N. L\"{u}tzgendorf;][]{Gonzalez2025}.
A dusty nearby source, \textit{Hyde}, was identified early from sub-mm data and later spectroscopically confirmed at the same redshift as \textit{Jekyll} \citep{Jekyll2018,Schreiber_2018,Schreiber_2021}. \textit{Hyde} is extremely dusty and vigorously star-forming ($\mathrm{SFR}_{\mathrm{IR}}\sim100\, M_\odot\, \mathrm{yr^{-1}}$), but the depth and spatial resolution of JWST/NIRCam now allow a robust deblending of its emission from that of \textit{Jekyll} (Section \ref{sec:photometry}).

The second target is SXDS-27434 at $z=4.01$, which was identified as the first massive QG at $z\gtrsim 4$ \citep{Tanaka_2019, Valentino2020}. Keck/MOSFIRE $K$-band spectrum shows strong Balmer absorption features, and recently \textit{DeepDive} targeted with G235M/F170LP medium-resolution slit spectroscopy.

 \subsection{Environmental Density Estimates}
 \label{sec:overdensity}
\begin{figure*}[tb]
    \plotone{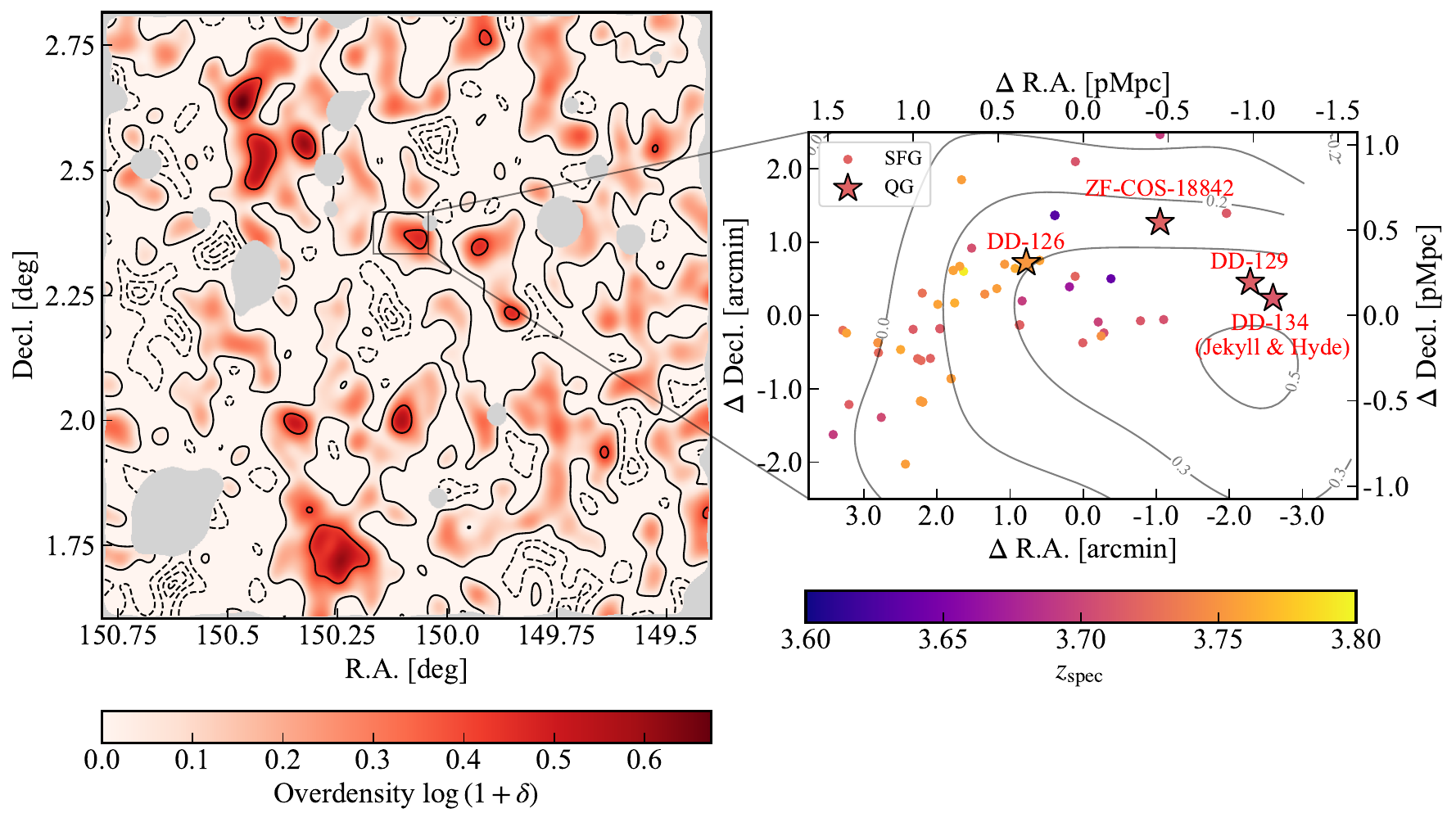}
    \caption{Distribution of massive ($M_\mathrm{*}>10^{10}\,M_\odot$) galaxies at $3.5<z_\mathrm{phot}<3.9$ in the entire COSMOS field. The red color shows the density excess of the massive galaxy candidates smoothed over $500\,\mathrm{pkpc}$ using Gaussian kernel density estimation. The black solid and dashed contour indicates the positive and negative overdensity value with 0.4 dex, respectively. The right panel shows locations of galaxies which are spectroscopically confirmed with JWST around the overdensity. The stars are QGs within the overdensity, and the points are SFGs which are spectroscopically confirmed at $3.6<z_\mathrm{spec}<3.8$ \citep[from the Dawn JWST Archive and COSMOS spec-z catalog;][]{Khostovan2026}. The gray contour corresponds to the overdensity value, and the QGs are located in the same overdensity.
    \label{fig:dens}}
\end{figure*}

To characterize the environmental density and the large-scale structure surrounding the targets, wide-field surveys are essential. We search for a local density enhancement around \textit{Jekyll} in the COSMOS field \citep{Scoville2007} and around SXDS-27434 in the Subaru/XMM-Newton Deep Survey \citep[SXDS;][]{Furusawa2008}. These surveys provide multiwavelength photometric data to probe a large sample of galaxies over a wide area.
To quantify the overdensity of galaxies, we utilize both massive SFGs and massive QGs from our photo-$z$ \citep{Tanaka_2015} based on the COSMOS2020 \textsc{Classic} catalog \citep{Weaver_2022} and the SXDS catalog \citep{Kubo2018}. The stellar mass thresholds are determined by the mass completeness of both QGs and SFGs, set at $M_*>10^{10}\,M_\odot$ for the COSMOS2020 catalog \citep{Weaver_2023} and $M_*>10^{10.5}\,M_\odot$ for the SXDS catalog \citep{Tanaka2024}.
We perform a kernel density estimation using a Gaussian kernel with a radius of $500\,\mathrm{pkpc}$\footnote{Hereafter, the prefix `p' indicates proper distances (e.g., pMpc, pkpc).}, guided by recent works reporting overdensities and proto-clusters around massive QGs at $z\gtrsim3$ on similar scales \citep{Chiang2013,McConachie2022, Ito2023, Tanaka2024}. The fixed-aperture method has been shown to be appropriate for tracing the super-halo scale environment \citep{Muldrew2012}. The strength of the overdensity is defined by following equation:
\begin{equation}
    \delta = \frac{n-\bar{n}}{\bar{n}},
\end{equation}
where $n$ is the galaxy surface number density at the location of each galaxy, and $\bar{n}$ is the average number density of the entire region. For this calculation, we consider redshift slices of $dz = 0.2$ (corresponding to $\sim 60\,\mathrm{pMpc}$). The redshift width is based on $\sim 1.5\sigma$ uncertainty between our photometric redshifts and the spectroscopic redshifts \citep{Ito_2022, Tanaka2024}, striking a compromise between contamination and completeness. The following results are largely insensitive to small changes ($1$--$2\sigma$) in the redshift range. In order to correct the overdensity mis-estimates by residual boundary effects, especially close to bright stars masked in the COSMOS2020 catalog (from the Suprime-Cam images), we apply the correction described in \cite{Chartab2020}. Specifically, random points that show the observed area are generated based on the masks, and the correction factor at each galaxy location is computed based on the number of surrounding random points. Since \textit{Jekyll \& Hyde} are located in the masked region based on the HSC images, we do not use that mask. For the SXDS field, we made random points based on all galaxies inside the catalog. If the correction factor exceeds a factor of $1.67$ (corresponding to a $1\sigma$ deviation from the mean value), we ignore the region when computing the average density of galaxies. The average number density of massive galaxies is $4.3\times 10^{-4}\,\mathrm{cMpc^{-3}}$\footnote{The prefix `c' indicates comoving distances.} at $3.5<z_\mathrm{phot}<3.9$ in COSMOS and $6.6\times 10^{-5}\,\mathrm{cMpc^{-3}}$ at $3.8<z_\mathrm{phot}<4.2$ in SXDS.

Figure \ref{fig:dens} shows the distribution of galaxy overdensities above the average field value at $3.5<z_\mathrm{phot}<3.9$ in the COSMOS field. 
Four QGs, including \textit{Jekyll}, are located in the same region, revealing a marginally significant overdensity ($\delta \sim 1.6$). 
The spatial scale of the overdensity around \textit{Jekyll} is $\sim 600\,\mathrm{pkpc}$ (average proper distance of the QGs), which corresponds to the typical size of a proto-cluster \citep{Overzier_2016,Chiang_2017,Lovell2018}.
The photometric association of QGs around \textit{Jekyll \& Hyde} is spectroscopically confirmed with JWST/NIRSpec from the literature. In fact, \citet{DeepDive_2025} and \citet{Nanayakkara2025} reported robust spectroscopic redshifts for three QGs at $z_{\rm spec}=3.71-3.75$ (Table \ref{tab:morp}; DD-126 and DD-129 in \citealt{DeepDive_2025} and ZF-COS-18842 in \citealt{Nanayakkara2025}), consistent with that of \textit{Jekyll} (DD-134). We also plot the spectroscopically confirmed galaxies (SFGs in the right panel of Figure \ref{fig:dens}) at $3.6<z_\mathrm{spec}<3.8$ from the Dawn JWST Archive (DJA) and the COSMOS spec-$z$ catalog \citep{Khostovan2026}, which also shows concentration of line emitters (low-mass SFGs). We note that these galaxies are not located at the peak of the overdensity. Although the spatial offset could hint at a tracer-dependent environmental structure (particularly between low-mass SFGs and massive galaxies), it is difficult to interpret because the mass limits are different for different spectroscopic observations and also different from the photometric data. 
In addition, statistical correction for the masked area due to the bright star close to \textit{Jekyll \& Hyde} may also affect the location of the overdensity peak.

\begin{figure}[tb]
    \plotone{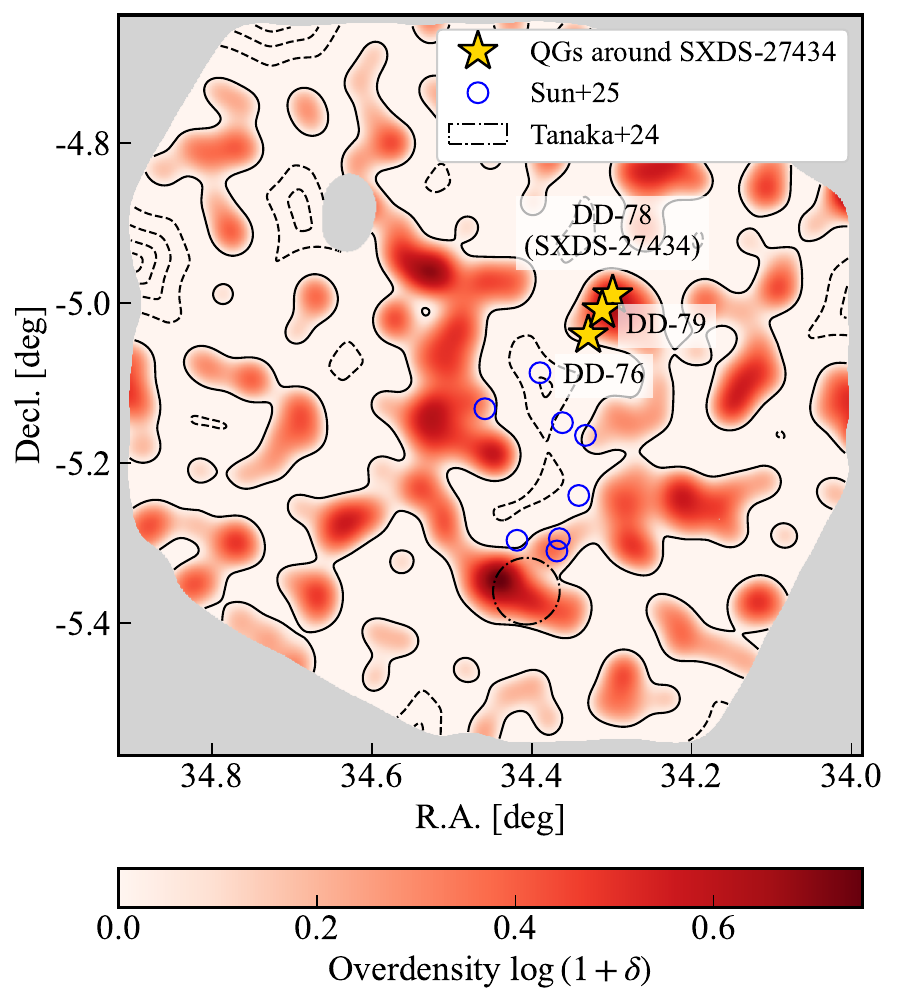}
    \caption{Distribution of massive ($M_\mathrm{*}>10^{10.5}\,M_\odot$) galaxies at $3.8<z_\mathrm{phot}<4.2$ in the entire SXDS field. The red color and the contour shows the density excess of the massive galaxy candidates smoothed over $500\,\mathrm{pkpc}$ using Gaussian kernel density estimation (as in the left panel of Figure \ref{fig:dens}). The yellow stars are QGs observed with \textit{DeepDive} within the overdensity. The black and blue circles are structures confirmed so far at $z\sim 4$ \citep{Tanaka2024, Sun2025}.
    \label{fig:densz4}}
\end{figure}

Figure \ref{fig:densz4} shows the distribution of galaxy overdensities at $3.8<z_\mathrm{phot}<4.2$ in the SXDS field. SXDS-27434 is also located within the large-scale overdensity ($\delta \sim 3.0$). Spectroscopic conformations of the proto-clusters or overdensities with QGs are reported at $z\sim 4.0$ in this field \citep{Tanaka2024, Sun2025}, and this suggests the existence of large-scale structure. The newly discovered overdensity is located at the north side of this structure, and two QGs are spectroscopically confirmed with JWST/NIRSpec from \textit{DeepDive} at $z_{\rm spec}=3.99-4.01$ (DD-76 and DD-79 in \citealt{DeepDive_2025}), which is consistent with that of SXDS-27434 (DD-78; Table \ref{tab:morp}). We include the spectroscopic data in our analysis later on this work.

\begin{deluxetable*}{cccccccc}
\digitalasset
\tablewidth{0pt}
\tablecaption{Coordinates, spectroscopic and photometric redshifts of QGs in the overdensities\label{tab:morp}}
\tablehead{
\colhead{Name} & \colhead{R.A.} & \colhead{Decl.} & \colhead{$\theta_{\mathrm{sep}}$} & \colhead{$z_\mathrm{spec}$} & \colhead{$z_\mathrm{phot}$} & \colhead{$z_\mathrm{phot}$} & \colhead{$z_\mathrm{phot}$} \\
 &&&[pkpc]&& [Our photo-$z$] & [COSMOS2025] & [PRIMER]
}
\startdata
     {DD-134} & $\mathrm{10^h00^m14^s.750}$ & $\mathrm{+02^\circ 22' 43''.358}$ & -- & $3.7120$ & $3.52^{+0.03}_{-0.06}$ & $0.98^{+0.03}_{-0.01}$ & $3.60^{+0.08}_{-0.07}$\\
     {DD-129} & $\mathrm{10^h00^m16^s.002}$ & $\mathrm{+02^\circ 22' 56''.504}$ & 146 & $3.7157$ & $3.45^{+0.41}_{-0.43}$ & $3.58^{+0.01}_{-0.08}$ & -- \\
     {DD-126} & $\mathrm{10^h00^m28^s.266}$ & $\mathrm{+02^\circ 23' 12''.686}$ & 1299 & $3.7511$ & $3.82^{+0.09}_{-0.18}$ & $3.96^{+0.08}_{-0.13}$ & $3.65^{+0.11}_{-0.10}$ \\
     {ZF-COS-18842} & $\mathrm{10^h00^m20^s.941}$ & $\mathrm{+02^\circ 23' 45''.748}$ & 800 & $3.7186$ & -- & $3.25^{+0.08}_{-0.02}$ & $3.53^{+0.06}_{-0.16}$ \\
     \hline
     {DD-78} & $\mathrm{02^h17^m11^s.690}$ & $\mathrm{-04^\circ 59' 23''.532}$ & -- & $4.0117$ & $3.97^{+0.10}_{-0.09}$ & -- & -- \\
     {DD-76} & $\mathrm{02^h17^m19^s.046}$ & $\mathrm{-05^\circ 02' 23''.640}$ & 1467 & $4.0125$ & $3.45^{+0.16}_{-0.15}$ & -- & -- \\
     {DD-79} & $\mathrm{02^h17^m14^s.866}$ & $\mathrm{-05^\circ 00' 31''.572}$ & 577 & $3.9933$ & $4.12^{+0.04}_{-0.05}$ & -- & -- \\
\enddata
\tablecomments{DD-134 is \textit{Jekyll}, and DD-78 is SXDS-27434. The object name is from \citet{DeepDive_2025} and  \citet{Nanayakkara2025}. $\theta_{\mathrm{sep}}$ is the projected physical distance between \textit{Jekyll} and the other members inside the COSMOS overdensity. For DD-76 and DD-79, $\theta_{\mathrm{sep}}$ is distance from DD-78. DD-134 and DD-129 are inside bright star masks in COSMOS2020 and COSMOS2025 (based on the HSC images).}
\end{deluxetable*}

\subsection{Overdensity Identification} \label{sec:cluster}
We also estimate the galaxy overdensity around \textit{Jekyll \& Hyde} using the COSMOS2025 catalog \citep{Shuntov_2025} and the PRIMER-COSMOS catalog, but there is no significant overdensity that includes all 4 QGs. For PRIMER-COSMOS, this may be due to the small field coverage and the problem of boundary corrections (\textit{Jekyll \& Hyde} is located near the boundary of PRIMER-COSMOS, and DD-129 is not inside the field). For the COSMOS2025 catalog, the differences between spectroscopic redshifts and photometric redshifts are large (especially for high-redshift QGs such as \textit{Jekyll}; $z_\mathrm{phot}=0.9756$), although the field is wider. Table \ref{tab:morp} shows the photometric redshifts of member galaxies from our photometric redshift catalog, the COSMOS2025 catalog, and the PRIMER-COSMOS catalog. Due to contamination from bright stars, especially for \textit{Jekyll} and DD-129, estimating photometric redshifts becomes more difficult than for the other members. In addition, \textit{Jekyll}'s photometry is contaminated from \textit{Hyde}. We require higher spatial resolution images for both the optical and NIR, and JWST observation becomes crucial to obtain more accurate photometric redshifts. This highlights that identifying QG overdensities at high redshift is challenging, and that confirming candidate QGs with spectroscopic follow-up is crucial. 

\section{JWST Observations and Data Analysis} \label{sec:obs}

\subsection{Photometry Extraction and Deblending}\label{sec:photometry}

\begin{figure}[tb]
    \plotone{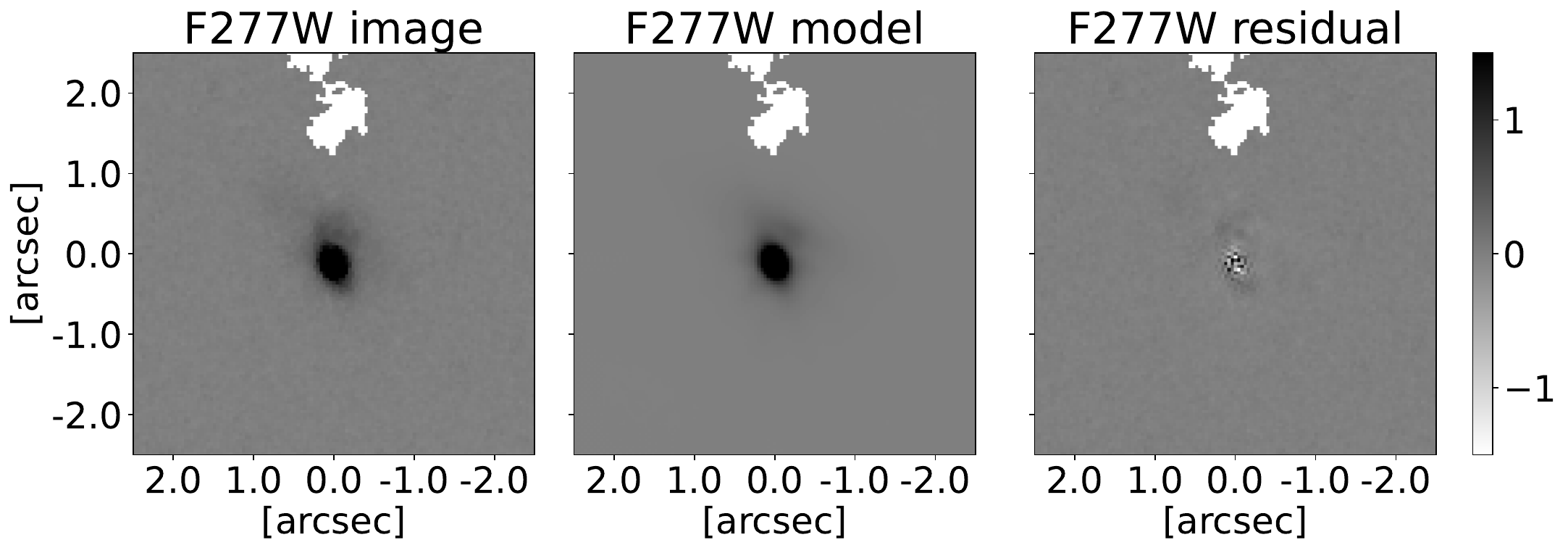}
    \caption{Observed NIRCam F277W science images, S\'{e}rsic model images, and residual images of \textit{Jekyll} (DD-134, $5'' \times 5''$) from left to right. The white regions show the masked regions to avoid the contamination of nearby galaxies.}
    \label{fig:image}
\end{figure}

\begin{figure}[tb]
    \plotone{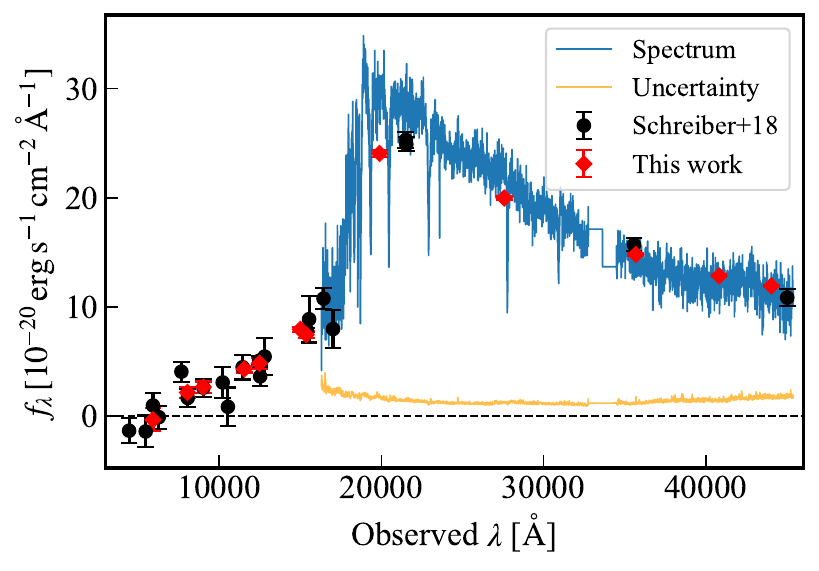}
    \caption{Comparison between \textit{Jekyll}'s photometry from \texttt{pysersic} (the red points) and in \cite{Jekyll2018} (the black points). The blue line shows the medium resolution spectrum of \textit{Jekyll} from \textit{DeepDive} \citep[DD-134;][]{DeepDive_2025}, and the yellow line is the spectrum uncertainty.\label{fig:QG1_photo}}
\end{figure}

In photometric catalogs such as PRIMER-COSMOS, \textit{Jekyll \& Hyde} suffers from blending because the pair is detected as a single source.
Given the proximity of \textit{Jekyll \& Hyde}, we model their surface brightness in NIRCam images with \cite{Sersic_1968} profile to deblend their emission and re-extract their photometry. We simultaneously fit the emission of both galaxies in all available HST (F606W, F814W, F125W, and F160W) and JWST/NIRCam (F090W, F115W, F150W, F200W, F277W, F356W, F410M, and F444W) from PRIMER using the Bayesian code \texttt{pysersic} \citep{Pasha2023}. We impose a smooth, continuous variation of the morphological parameters as a function of wavelength. For this exercise, we use the images taken from DJA (v7\footnote{\url{https://dawn-cph.github.io/dja/imaging/v7/}}, \citealt{Valentino_2023}). Nearby and possibly contaminating objects are masked using \texttt{sep} \citep{Barbary_2016, 1996A&AS..117..393B}. Figure \ref{fig:image} shows the F277W band images and S\'{e}rsic model images of \textit{Jekyll \& Hyde} from \texttt{pysersic}.
The shape of the SED from the total flux densities is consistent within 1$\sigma$ with the original photometry in \cite{Jekyll2018} in the overlapping wavelength ranges (Figure \ref{fig:QG1_photo}). In the rest of the analysis, we will use our updated HST and JWST photometries for \textit{Jekyll \& Hyde}. For the rest of the sample, we adopt the PSF-matched photometry from the \textit{DeepDive} catalog \citep{DeepDive_2025, Hamadouche2026}. For this case, we use the images with point spread functions (PSFs) matched to the reddest band (F444W) following \cite{Weaver_2024} (see \citealt{DeepDive_2025} for a full description). 

\subsection{Spectral Analysis}

\begin{figure*}[tb] 
    \plotone{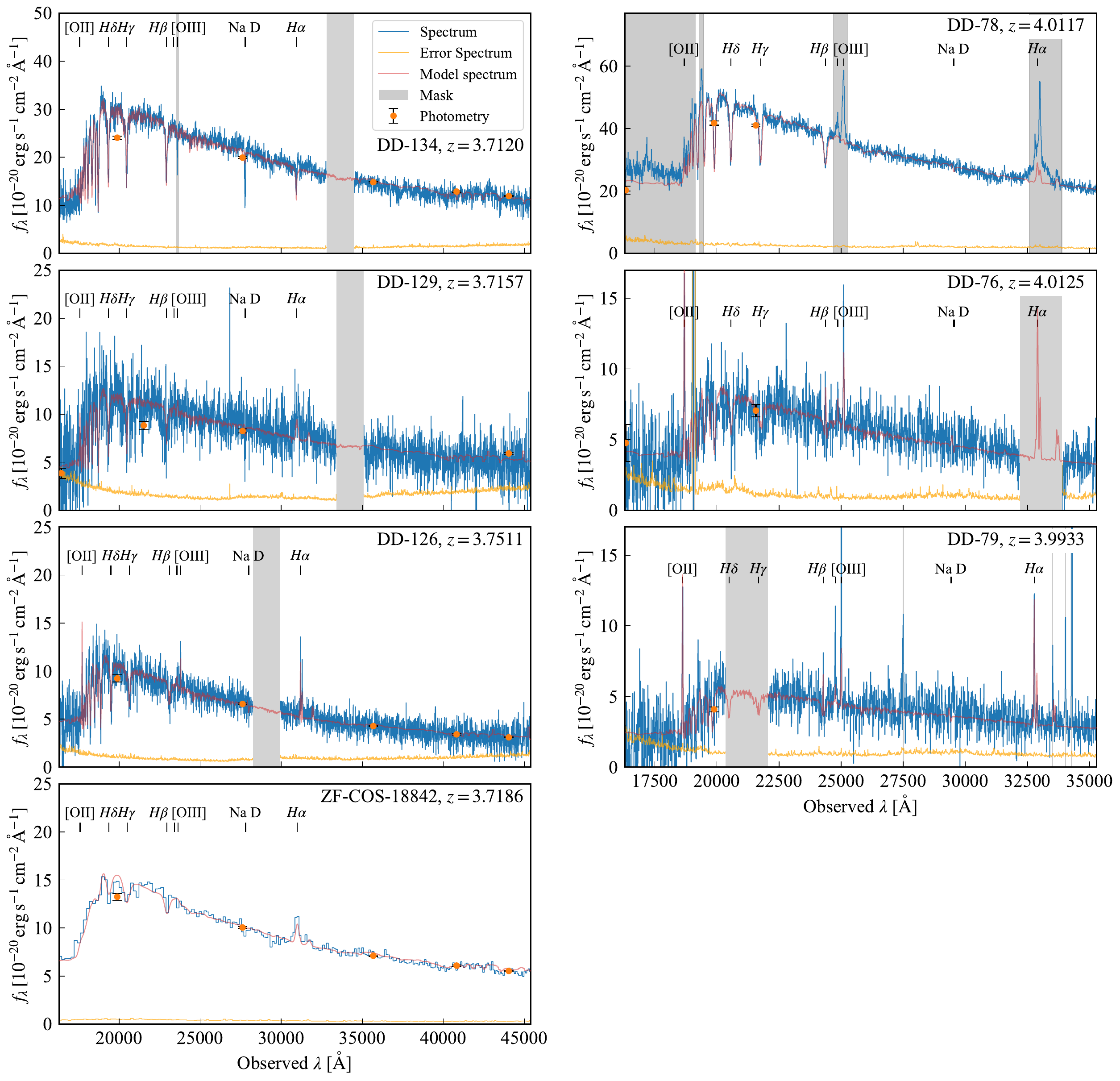}
    \caption{Observed spectra of the overdensity members with JWST (the blue line) and best-fitted SED from \texttt{prospector} (the red line; see Section \ref{sec:prospector}). The yellow lines show the spectrum uncertainty, and the orange points show the photometry obtained from PRIMER-COSMOS and COSMOS-Web for QGs around \textit{Jekyll} (DD-134), and \textit{DeepDive} direct images or WFCAM from UKIDSS \citep{Lawrence2007} for QGs around SXDS-27434 (DD-78). The gray shaded region is masked in the \texttt{prospector} fitting.}
    \label{fig:Spec}
\end{figure*}

As mentioned in Section \ref{sec:overdensity}, NIRSpec G235M/F170LP spectra of six QG members are available from the literature. Spectra of DD-134 (\textit{Jekyll}), DD-129, DD-126, DD-79, DD-78 (SXDS-27434), and DD-76 are taken from the \textit{DeepDive} program \citep{DeepDive_2025}. The NIRSpec PRISM/CLEAR spectrum of the other QG (ZF-COS-18842), reported in \citet{Nanayakkara2025}, is taken from DJA (v4; \citealt{zenodo, Valentino_2025}). All spectra are reduced with \texttt{msaexp} \citep{deGraaff_2025A, Heintz_2025} and are scaled to match with the total photometry using 0th to 2nd-order polynomials according to the number of broadband photometry within the spectral wavelength range (the spectrum of DD-134 is scaled to match with the photometry extracted in Section \ref{sec:photometry}).\\

Figure \ref{fig:Spec} shows the spectra of the QG members. The medium resolution grating spectra of QGs around DD-134 show strong Balmer absorption features. Their spectroscopic redshifts are all similar to \textit{Jekyll \& Hyde} (Table \ref{tab:morp}). The fourth member (ZF-COS-18842) also has the same redshift value. The significant Balmer breaks and Balmer absorption lines indicate that they experienced rapid quenching in the recent past, and they currently have no significant ongoing star formation. We note that the spectrum from \textit{Jekyll} shows an absorption feature at the same wavelength as [O\,{\footnotesize III}]$\lambda5007$, but this is a calibration effect due to contamination from \textit{Hyde} (IFU spectrum of \textit{Jekyll} described in \cite{Gonzalez2025} shows the emission line). We masked out the [O\,{\footnotesize III}] line as well as the spectral gap due to the detector gap from the SED fitting analysis in the following Sections.

In contrast, the spectra of the member galaxies inside SXDS overdensity (DD-76, DD-78, and DD-79) show the narrow and/or broad emission lines. DD-78 (SXDS-27434) newly confirms clear Balmer absorption lines as well as the broad H$\alpha$ emission and strong [O\,{\footnotesize III}] emission lines ($EW_\mathrm{[OIII]} = 13.3\,$\AA). This galaxy is also X-ray detected \citep{DeepDive_2025}. These facts clearly suggest that the galaxy has a powerful AGN at the center (Section \ref{subsec:AGN}). The other two galaxies only show the narrow H$\alpha$ and/or [O\,{\footnotesize III}] emission lines but still show the Balmer absorption lines and break around rest-frame 4000\,\AA. 

\subsection{Spectro-photometric Fitting} \label{sec:prospector}
We modeled the combined photometry and spectra (Figure \ref{fig:Spec}) of all QGs in the overdensity with \texttt{prospector} \citep{Johnson_2021}. 
We use the Flexible Stellar Population Synthesis (FSPS) model \citep{Conroy_2009,Conroy_2010} to infer physical properties of the galaxies. We further assume the \cite{Chabrier_2003} IMF, solar metallicities, the \cite{Madau_1995} IGM absorption model, and the attenuation curve in \cite{Calzetti_2000}. For the dust optical depth, we assume a top-hat prior between $0<\tau_V<6$. We fix the redshift to the spectroscopic estimate. The model includes nebular emission (with a fixed gas ionization parameter $U=0.001$ as in \citealp{carnall_2023}). We assume a non-parametric SFH model, which allows the SFR in each age bin to vary. Specifically, the two youngest age bins are fixed at 31.6\,Myr and 100\,Myr, and the remaining bins are defined in 0.15 dex intervals up to the age of the Universe at the observed redshift. Here we adopt a log-uniform prior to constrain the stellar masses of each age-bin \citep[the ``log-mass prior''][]{Leja_2017,Leja_2019}. We run the Markov Chain Monte Carlo (MCMC) code \texttt{emcee} \citep{2013PASP..125..306F} to infer the best-fitting parameters and their associated errors. We note that the broad emission line and AGN UV continuum were not modeled in the stellar population synthesis, and we masked them out in the DD-78 spectrum from the SED fitting analysis. In addition, the DD-79 spectrum has contamination from a background galaxy, and we masked out related bright [O\,{\footnotesize III}], H$\beta$, and H$\alpha$ emission lines and rest-frame UV photometry corresponding to the $Ly\alpha$ emission line. We mask a small portion of the DD-76 spectrum affected by bad pixels.

\section{Results} \label{sec:Results}
\subsection{Physical properties and star formation histories} \label{subsec:sfhs}
\begin{figure}[tb]
    \plotone{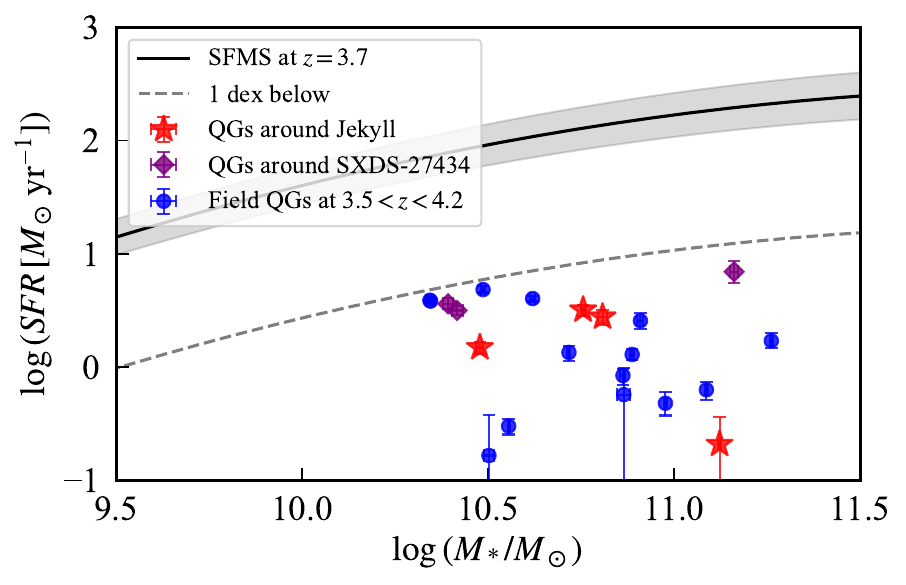}
    \caption{SFR plotted against stellar mass. The black line shows the star formation main sequence at $z\sim 3.7$ from \cite{Popesso2023}, and the gray dashed line is the same sequence shifted by -1 dex downwards. The red stars are QGs around \textit{Jekyll} at $z\sim 3.7$, and the purple diamonds are QGs around SXDS-27434 at $z\sim 4.0$. The blue points are massive QGs at $3.5<z<4.2$ from \citet{DeepDive_2025} for field sample (see Section \ref{subsec:sfhs}).
    \label{fig:SFMS}}
\end{figure}

Prospector SED fitting results indicate that member galaxies are likely quiescent. Figure \ref{fig:SFMS} shows the comparison between stellar masses and star formation rates averaged over 31.6\,Myr (in the youngest bin) of all QGs in the overdensity. All members are located more than 1 dex below the star-forming main-sequence \citep[SFMS;][]{Popesso2023}, which confirms that the members are all quenched, supporting the selection of QGs and stellar mass estimates in \cite{DeepDive_2025}. Even with the posterior width being underestimated due to model misspecification \citep{Jespersen2025_opticalIR}, this large distance assures that the selected galaxies represent a fully quenched sample.

\begin{figure*}[tb]
    \centering
    \begin{minipage}{0.49\textwidth}
        \centering
        \includegraphics[width=0.95\textwidth]{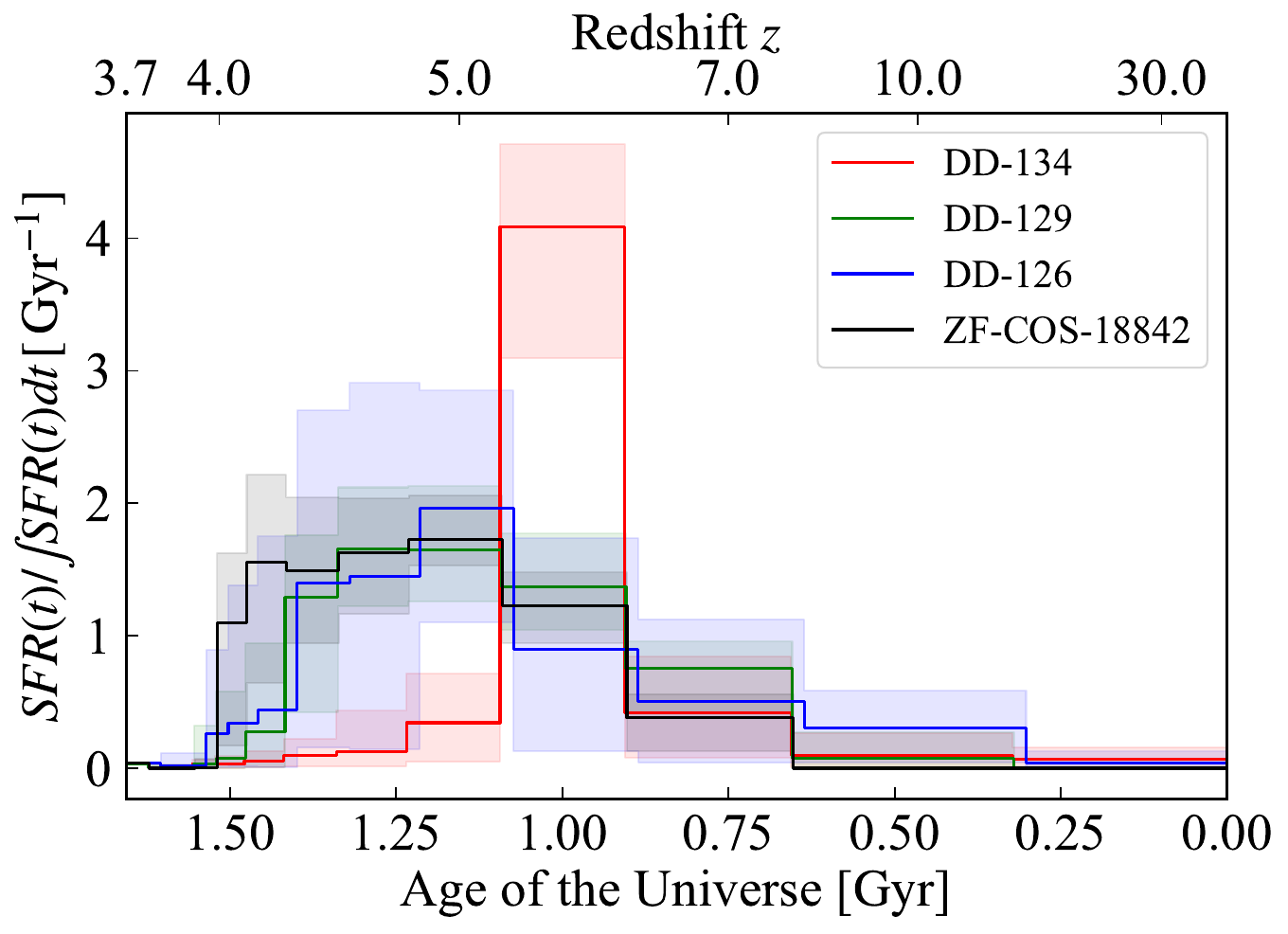}
    \end{minipage}
    \begin{minipage}{0.49\textwidth}
        \centering
        \includegraphics[width=0.95\textwidth]{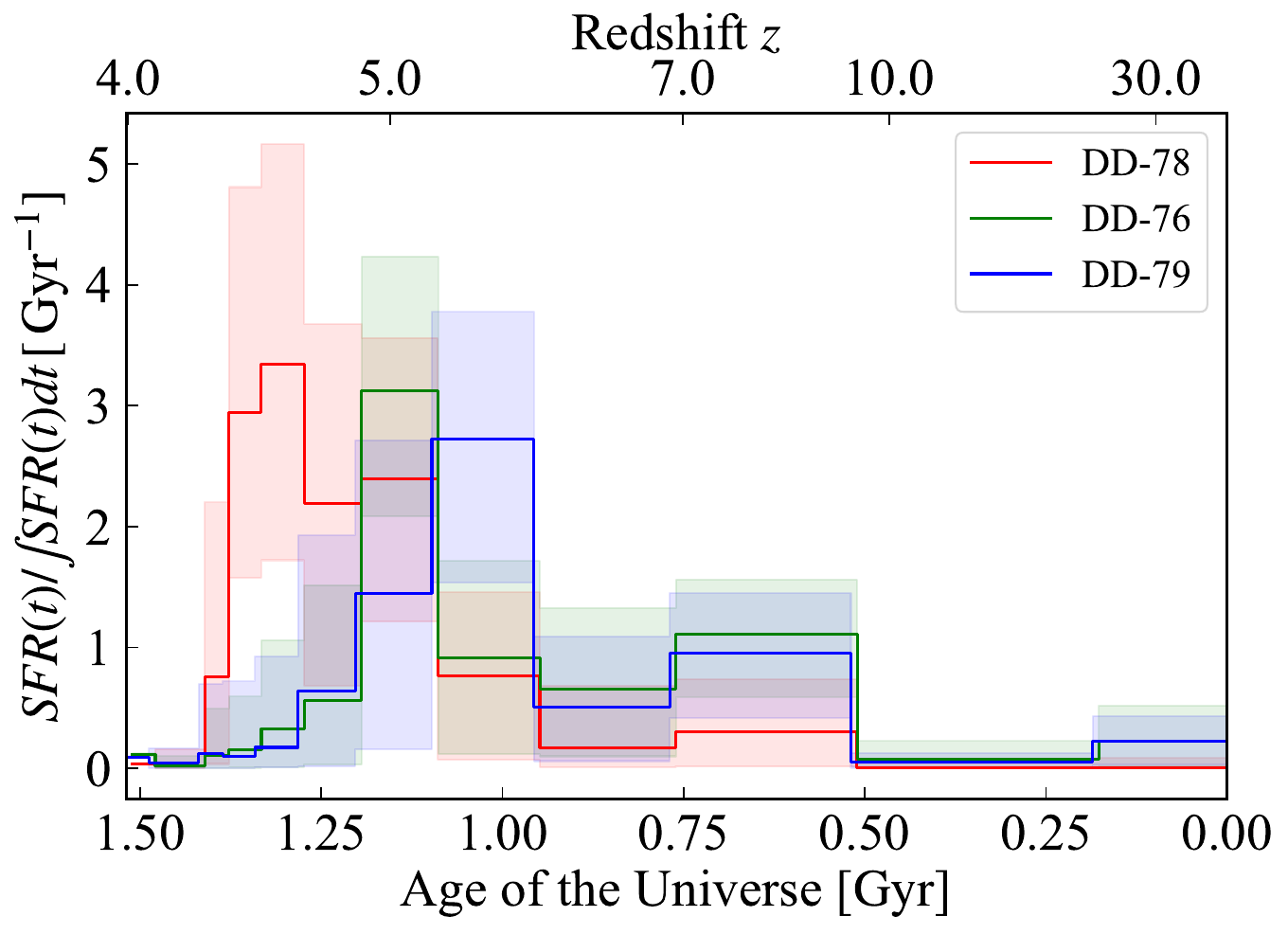}
    \end{minipage}
    \caption{Left Panel: Normalized star formation histories of four QGs at $z\sim 3.7$ around \textit{Jekyll} (DD-134) inferred from \texttt{prospector}. The lines are the best-fitted SFHs, and the shaded regions are the 68\% interval range. Right Panel: Same as the left panel but for QGs at $z\sim 4.0$ in SXDS.}
    \label{fig:SFH}
\end{figure*}

The star-formation history of the members also shows recent quenching. The left panel of Figure \ref{fig:SFH} shows the normalized SFH of each member galaxy around \textit{Jekyll}. \textit{Jekyll} experienced a starburst phase at $z\sim 5.2$ and rapidly quenched about $500\,\rm Myr$ prior to the time of observation. The other three QGs were still forming stars down to $z\sim 4$ and then quenched about $100\,\rm Myr$ prior to the time of observation. 
The SFHs inferred from massive QGs around SXDS-27434 also show a similar trend (the right panel of Figure \ref{fig:SFH}). DD-78 experienced a starburst phase at $z=4.5$ and rapidly quenched about $100\,\rm Myr$ prior to the time of observation, which is similar to \textit{Jekyll}, particularly in terms of the rapid timescale of the starburst and its subsequent quenching. DD-76 and DD-79 were quenched about $200\,\rm Myr$ prior to the time of observation. 
We note that the physical properties and star formation history of DD-134 (\textit{Jekyll}) are consistent with the integrated IFU spectrum results \citep{Gonzalez2025}. The physical properties of DD-78 are also consistent with the ground-based observations \citep{Valentino2020}. In addition, DD-129 has an ALMA Band 6 continuum detection \citep{Suzuki_2022}, and its inferred $SFR_\mathrm{IR}$ is marginally consistent with $SFR_\mathrm{SED}$.

Interestingly, member galaxies inside the COSMOS and SXDS field overdensities have similar formation and quenching redshifts ($z_\mathrm{form}, z_\mathrm{quench}$). $z_\mathrm{form}$ and $z_\mathrm{quench}$ are defined as the redshifts at which the galaxy formed 50\% and 90\% of their total stellar mass at the observed epoch, respectively. Table \ref{tab:prop} summarizes their physical properties. The quenching redshifts are slightly different, especially between the most massive galaxy and the others, but the members seem to have very similar $z_\mathrm{form}$ and $z_\mathrm{quench}$ within each overdensity. Although limited by sample statistics, this hints at similar SFHs of the member QGs.

\begin{figure}[tb]
    \plotone{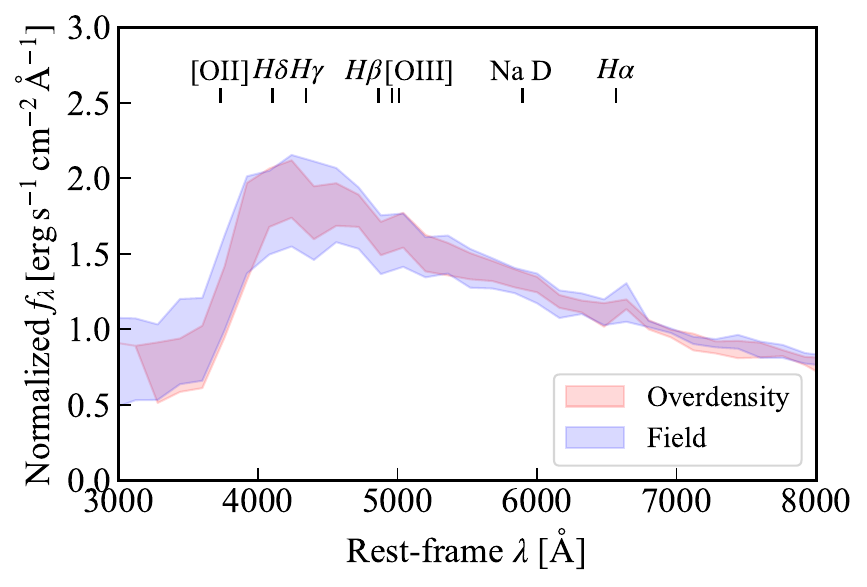}
    \caption{Scatter of the spectra from QGs in the overdensities (the red shaded region) and field QGs in \cite{DeepDive_2025} (the blue shaded region). The shaded regions represent $1\sigma$ scatters of the spectra for the QG sample. The spectra are normalized between rest-frame 6800 and 7000\,\AA.
    \label{fig:stack}}
\end{figure}

The SFHs similarity is supported by their agreement in spectral shapes; Figure \ref{fig:stack} shows the spectral scatter of QGs inside the overdensities and QGs in the field. For comparison, we construct a sample of field QGs based on $Dn(4000)$ \citep{Balogh1999}, $UVJ$ color \citep{Williams2009}, or $sSFR$ criteria from \citet{DeepDive_2025}. Fourteen QGs are confirmed by \textit{DeepDive} and JWST archival sample (from DJA) at $3.5<z_\mathrm{spec}<4.2$, which are not located in the same overdensity region\footnote{DD-82 is also classified as a $UVJ$-selected QG in \citet{DeepDive_2025}, but the SED fitting result using \texttt{prospector} indicates the high SFR (nearby SFMS). In the following analysis, we do not include the galaxy as a field sample.}. We also include the inferred physical properties of the field QGs in the bottom of Table \ref{tab:prop}. The shaded region in Figure \ref{fig:stack} shows the 68\% scatter of the spectra for each of the overdensity and field QG sample. The spectra are normalized between rest-frame 6800 and 7000\,\AA. In this comparison, we bin the spectra into $\sim150$\,\AA\ intervals to minimize the effects of varying signal-to-noise ratios between the spectra. The figure indicates that the QGs inside the overdensities exhibit a smaller spectral scatter in Balmer break than the field QGs. To calculate the significance of this scatter, we randomly selected seven QGs from both the field and overdensity samples with replacement, and then perturbed each spectrum based on its observational uncertainty. Finally, the spectral scatter among the seven spectra was calculated in the rest-frame 4000--4100\,\AA\ range. This bootstrap simulation reveals that the scatter is $1.41 \pm 0.18$ times larger in field QGs than in overdensity QGs, with a significance level exceeding $2\sigma$. This scatter can be interpreted as the scatter in the stellar ages of the galaxies, suggesting that the QGs inside the overdensities indeed follow similar SFHs, and they experienced intense star formation and subsequent quenching about the same time. We note that the stellar mass and redshift distribution of field QGs is also consistent with the QGs in the overdensities (Figure \ref{fig:SFMS}), and there is no age dependence with stellar mass inside the sample. We further compare the formation times and quenching times in Section \ref{sec:Discussion}.

\begin{deluxetable*}{ccccccc}
\digitalasset
\tablewidth{0pt}
\tablecaption{Inferred physical properties of the QGs in the overdensities and in the field \label{tab:prop}}
\tablehead{
\colhead{Name} & \colhead{$z_\mathrm{spec}$} & \colhead{$\log{M_*/M_\odot}$} & \colhead{$SFR_\mathrm{SED}$} & \colhead{$A_V$} & \colhead{$z_\mathrm{form}$}  & \colhead{$z_\mathrm{quench}$} \\
 &&& [$M_\odot\,\rm yr^{-1}$] & [mag] & &
}
\startdata
     {DD-134} & 3.712 & $11.12\pm 0.01$ & $0.1^{+0.2}_{-0.1}$ & $0.22\pm 0.03$ & $5.72^{+0.15}_{-0.07}$ & $5.37^{+0.03}_{-0.08}$ \\
     {DD-129} & 3.7157 & $10.81\pm 0.02$ & $2.8\pm 0.5$ & $0.69^{+0.05}_{-0.06}$ & $5.15^{+0.18}_{-0.12}$ & $4.54^{+0.12}_{-0.10}$ \\
     {DD-126} & 3.7511 & $10.48\pm 0.01$ & $1.5\pm 0.2$ & $0.17\pm 0.07$ & $5.12^{+0.25}_{-0.20}$ & $4.51^{+0.15}_{-0.13}$ \\
     {ZF-COS-18842} & 3.7186 & $10.75\pm 0.01$ & $4.9\pm 0.8$ & $0.53^{+0.04}_{-0.06}$ & $4.87^{+0.22}_{-0.13}$ & $4.35^{+0.15}_{-0.07}$ \\
     \hline
     {DD-78} & 4.0117 & $11.16\pm 0.01$ & $6.9^{+1.7}_{-1.5}$ & $0.43^{+0.04}_{-0.06}$ & $4.85\pm 0.15$ & $4.39^{+0.05}_{-0.03}$ \\
     {DD-76} & 4.0125 & $10.39\pm 0.02$ & $3.6^{+0.5}_{-0.4}$ & $0.07^{+0.06}_{-0.05}$ & $5.66^{+0.50}_{-0.43}$ & $4.87^{+0.05}_{-0.18}$ \\
     {DD-79} & 3.9933 & $10.41\pm 0.02$ & $3.2^{+0.4}_{-0.3}$ & $0.41\pm 0.07$ & $5.59\pm 0.18$ & $4.83^{+0.17}_{-0.19}$ \\
     \hline\hline
     {DJA-65} & 4.129 & $10.49\pm 0.01$ & $4.8^{+0.4}_{-0.3}$ & $0.35\pm 0.04$ & $5.64^{+0.16}_{-0.15}$ & $5.14^{+0.05}_{-0.06}$ \\
     {DJA-71} & 4.1065 & $10.86\pm 0.01$ & $0.8\pm 0.2$ & $0.27^{+0.07}_{-0.05}$ & $6.07^{+0.44}_{-0.37}$ & $4.67^{+0.20}_{-0.17}$ \\
     {DD-80} & 3.9879 & $11.09\pm 0.01$ & $0.6\pm 0.2$ & $0.25^{+0.06}_{-0.03}$ & $6.29\pm 0.07$ & $5.70^{+0.15}_{-0.56}$ \\
     {DJA-91} & 3.9783 & $10.86\pm 0.02$ & $0.9^{+0.4}_{-0.3}$ & $1.45^{+0.09}_{-0.11}$ & $6.63^{+3.94}_{-0.94}$ & $4.66^{+0.17}_{-0.09}$ \\
     {DJA-92} & 3.9746 & $10.62\pm 0.01$ & $4.0\pm 0.3$ & $0.46\pm 0.02$ & $4.73^{+0.12}_{-0.07}$ & $4.35^{+0.06}_{-0.03}$ \\
     {DD-96} & 3.8676 & $10.98\pm 0.01$ & $0.5\pm 0.2$ & $0.02^{+0.03}_{-0.02}$ & $5.95^{+0.06}_{-0.05}$ & $5.24^{+0.31}_{-0.26}$ \\
     {DD-106} & 3.8307 & $11.26\pm 0.01$ & $1.7\pm 0.3$ & $0.05^{+0.07}_{-0.03}$ & $5.86^{+0.08}_{-0.19}$ & $5.08^{+0.29}_{-0.28}$ \\
     {DD-111} & 3.7976 & $10.91\pm 0.01$ & $2.6^{+0.5}_{-0.4}$ & $0.81^{+0.05}_{-0.06}$ & $4.69^{+0.11}_{-0.14}$ & $4.14^{+0.05}_{-0.06}$ \\
     {DD-115} & 3.7766 & $10.89\pm 0.01$ & $1.3\pm 0.2$ & $0.09^{+0.03}_{-0.02}$ & $4.43^{+0.02}_{-0.03}$ & $4.09\pm 0.02$ \\
     {DJA-134} & 3.7013 & $10.55\pm 0.01$ & $0.3\pm 0.1$ & $0.05\pm 0.03$ & $5.39^{+0.11}_{-0.15}$ & $4.68^{+0.11}_{-0.17}$ \\
     {DD-144} & 3.6507 & $10.34\pm 0.02$ & $3.9\pm 0.3$ & $0.03\pm 0.03$ & $4.52^{+0.06}_{-0.04}$ & $4.29^{+0.08}_{-0.11}$ \\
     {DJA-156} & 3.6053 & $10.72\pm 0.01$ & $1.3^{+0.3}_{-0.2}$ & $0.65\pm 0.04$ & $4.94^{+0.13}_{-0.07}$ & $4.52^{+0.09}_{-0.13}$ \\
     {DJA-163} & 3.5851 & $10.35\pm 0.01$ & $3.8\pm 0.2$ & $0.05\pm 0.04$ & $5.32^{+0.33}_{-0.16}$ & $4.15^{+0.22}_{-0.14}$ \\
     {DD-165} & 3.51 & $10.50\pm 0.02$ & $0.2^{+0.3}_{-0.2}$ & $0.34^{+0.15}_{-0.08}$ & $4.59^{+0.28}_{-0.25}$ & $4.05^{+0.14}_{-0.16}$ \\
     \hline
\enddata
\tablecomments{The object name is based on \citet{DeepDive_2025}  and \citet{Nanayakkara2025}. The first four galaxies is around \textit{Jekyll \& Hyde} (DD-134), the following three galaxies are around SXDS-27434 (DD-78), and the others are field sample selected in Section \ref{subsec:sfhs}. See Section \ref{subsec:sfhs} for the definition of $z_\mathrm{form}$ and $z_\mathrm{quench}$.}
\end{deluxetable*}

\subsection{Effects of AGNs} \label{subsec:AGN}
From the medium-resolution spectrum, the broad Balmer emission line is confirmed only from DD-78 (SXDS-27434), but some member galaxies show faint and narrow Balmer and [O\,{\footnotesize III}] emission lines. We attempt to classify them as either AGN or star formation origin based on the emission line intensity ratios. We utilize penalized Pixel Fitting algorithm \citep[\texttt{pPXF};][]{Cappellari2017,Cappellari2023} to fit the emission lines and continuum. Since the stellar component and emission lines can be modeled separately using FSPS templates and Gaussian line profiles, respectively, robust emission line fluxes can be measured by accounting for the underlying stellar absorption lines.
We fit the narrow emission lines with a single Gaussian component except DD-78, for which we add the broad Balmer components to H$\gamma$, H$\beta$, and H$\alpha$.

\begin{figure}[tb]
    \plotone{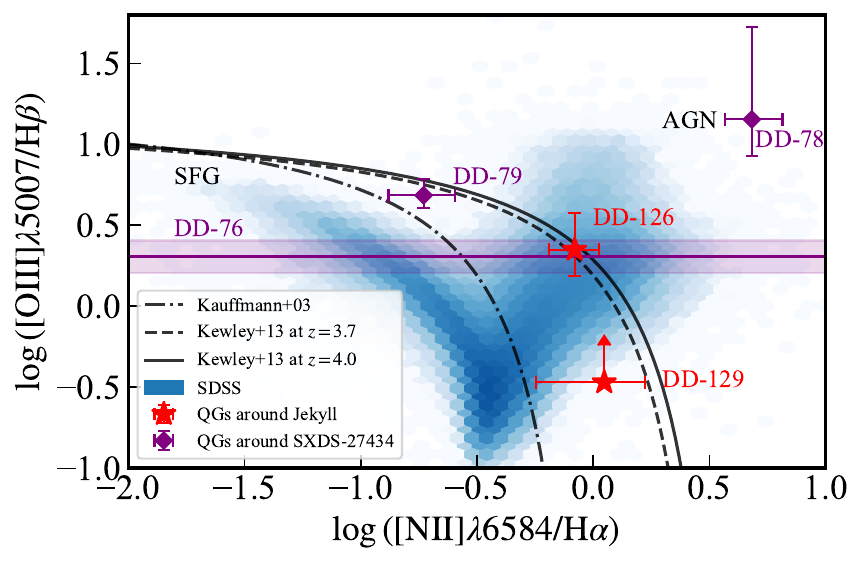}
    \caption{[O\,{\footnotesize III}]/H$\beta$ plotted against [N\,{\footnotesize II}]/H$\alpha$. The symbols show DD-126, DD-78, and DD-79 using JWST medium resolution spectra from \textit{DeepDive}. DD-129 shows $3\sigma$ lower limit of [O\,{\footnotesize III}]/H$\beta$ ratio based on its non-detection of H$\beta$. DD-76 shows only [O\,{\footnotesize III}]/H$\beta$ ratio due to the missing data around H$\alpha$ and [N\,{\footnotesize II}] emission lines. The black lines are the criteria to separate AGNs from star-forming galaxies from \cite{Kauffmann2003}, and these are extrapolated from \cite{Kewley2013} to $z=3.7$ and $4.0$. Blue shades indicate the galaxies in the local Universe from the MPA-JHU SDSS catalog \citep{Brinchmann2004, Kauffmann2003, Tremonti2004}. \label{fig:BPT}}
\end{figure}

The AGN classification is performed using the BPT diagram \citep{Baldwin1981}, which is reliable for metal-rich (massive) galaxies at high redshifts \citep{Hirschmann2019}. Figure \ref{fig:BPT} shows the AGN classification schemes from literature \citep{Kauffmann2003, Kewley2013} and observed line flux ratios for DD-129, DD-126, DD-78, and DD-79 in the [O\,{\footnotesize III}]/H$\beta$ vs. [N\,{\footnotesize II}]/H$\alpha$ diagram. Here, we use the narrow-line fluxes. We do not show DD-134 due to the non-detection of [O\,{\footnotesize III}], H$\beta$, and H$\alpha$ emission lines, and DD-76 due to the missing data around H$\alpha$ and [N\,{\footnotesize II}] emission lines. On the BPT diagnostic diagram, DD-78 is located in the AGN region, and DD-126 and DD-79 have a possibility of being classified as AGNs (or composite spectra). The H$\beta$ emission line is not detected in DD-129 but the lower limit is located in the composite region. This indicates that about half of the members are likely to host an AGN, a fraction that is consistent with the literature \citep[e.g.,][]{Bugiani2025, Skarbinski2025, Baker2025_FLAMINGO, Stevenson2026}. Interestingly, these AGN hosts tend to be recently quenched galaxies (i.e., the difference of $z_\mathrm{quench}$ and $z_\mathrm{spec}$ is small; $\sim 270\,\mathrm{Myr}$). We note that ZF-COS-18842 is not shown in the diagram due to its lower spectral resolution, but \cite{Kawinwanichakij2026} argue that it possibly has an AGN component based on its broader H$\alpha$ emission. The results suggest that the QGs inside the overdensities also experienced AGN feedback \citep[e.g.,][]{Man2021, Jin2024, Ito2025b}.

\section{Discussion: Implications of our findings for galaxy quenching} \label{sec:Discussion}

\subsection{Formation Epochs of the Members} \label{sec:formz}
\begin{figure*}[tb]
    \plotone{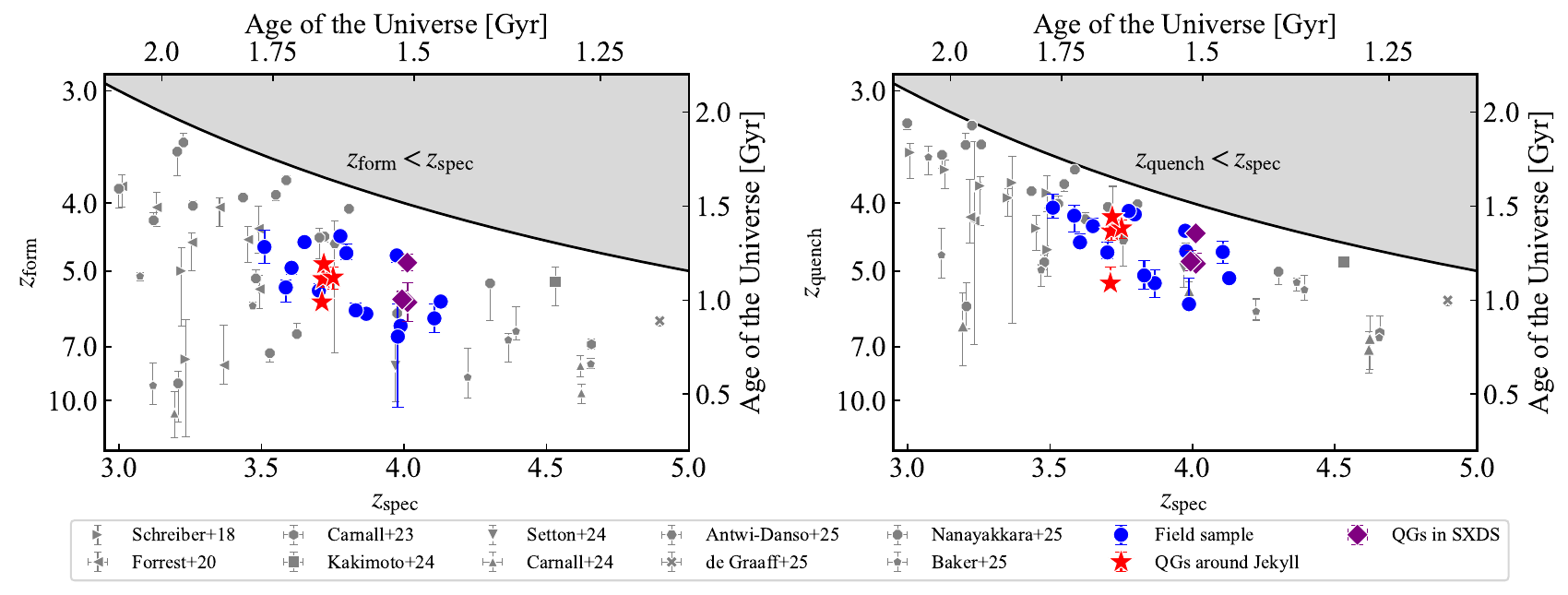}
    \caption{Left Panel: Formation redshift ($z_\mathrm{form}$) at which the galaxy formed 50\% of total stellar mass vs. observed redshift ($z_\mathrm{spec}$) of massive QGs confirmed thus far at $3.0<z_\mathrm{spec}<5.0$ \citep{Schreiber_2018,Forrest_2020,carnall_2023,Kakimoto2024,Carnall2024,Setton2024,AntwiDanso2025,deGraaff2025,Nanayakkara2025,Baker2025_FLAMINGO}. The red stars and purple diamonds are overdensity members around \textit{Jekyll} (DD-134) and SXDS-27434 (DD-78), respectively. The blue points are field QGs confirmed by JWST/NIRSpec medium resolution spectra at $3.5<z_\mathrm{spec}<4.2$ in \cite{DeepDive_2025}. The error bars are the uncertainty from the 68\% interval ranges. The black line represents the age of the Universe as a function of $z_\mathrm{spec}$ so that the gray region is outside of the prior of $z_\mathrm{form}$. Right Panel: Quenching redshift ($z_\mathrm{quench}$) at which the galaxy formed 90\% of total stellar mass vs. observed redshift ($z_\mathrm{spec}$) of massive QGs at $3.0<z_\mathrm{spec}<5.0$. We note that the definition of quenching time is different in some literature (using the time since the galaxy passed some quenching thresholds).}
    \label{fig:zform}
\end{figure*}

The tighter scatter in the SFHs of the QGs inside the overdensities becomes clearly evident when  comparing their formation ($z_\mathrm{form}$) and quenching ($z_\mathrm{quench}$) redshifts against those of field QGs.
The left panel of Figure \ref{fig:zform} compares the observed redshifts ($z_\mathrm{spec}$) and $z_\mathrm{form}$ of massive QGs ($M_*>10^{10}\,M_\odot$) spectroscopically confirmed so far at $3<z<5$. Compared to the diverse formation epochs of field QGs from the literature (e.g., $z_\mathrm{form} = 4.0$--$11.0$), the QGs in the overdensity regions exhibit relatively similar $z_\mathrm{form}$ and $z_\mathrm{quench}$, suggesting that they formed and quenched at about the same time. Furthermore, the blue points in Figure \ref{fig:zform} show the field QGs summarized in \citet{DeepDive_2025} at $3.5<z<4.2$ (14 QGs selected in Section \ref{subsec:sfhs}), indicating the diverse formation epochs ($z_\mathrm{form} = 4.4$--$6.6$). The right panel of Figure \ref{fig:zform} also shows that these field QGs have widely distributed quenching epochs ($z_\mathrm{quench} = 4.0$--$5.7$).

In order to quantify the similarity in formation and quenching epochs of the QGs within the overdensities, we perform a controlled statistical comparison against the 14 field QGs. Thanks to the comparable quality and wavelength coverage of the spectra and photometry from both JWST and ground-based telescopes between the field QGs and those in overdense regions, we can estimate the physical properties of these QGs using the same method summarized in Section \ref{sec:prospector}. We randomly drew subsets of 7 galaxies from the 14 field QGs and estimate the scatter in their formation times, $t_\mathrm{form}$ (the stellar ages corresponding to $z_\mathrm{form}$), via 50,000 Monte Carlo realizations. For each galaxy, we drew 50,000 samples of $t_\mathrm{form}$ directly from the MCMC posteriors of the SED fitting, and they are marginalized over the other parameters. In order to ensure a robust comparison against outliers, we evaluate the equivalent $1\sigma$ scatter based on the interquartile range (i.e., the difference between the 75th and 25th percentiles divided by 1.35). We note that ZF-COS-18842 is observed only with the PRISM mode and does not have the same spectral resolution as the other objects. We also compare the SED fitting results of \textit{Jekyll} from both DD-134 (the medium resolution spectrum from \textit{DeepDive}) and ZF-COS-20115 \citep[the PRISM spectrum from][]{Nanayakkara2025}, which inferred the consistent $z_\mathrm{form}$ and $z_\mathrm{quench}$ within $1\sigma$. In addition, some field QGs are not from \textit{DeepDive} original sample but from DJA, which have larger spectral coverage for the rest-frame UV side. We also perform the same SED fitting analysis using a consistent wavelength range with that of \textit{DeepDive}, finding consistent physical properties. In summary, our overdensity sample and field QGs are homogeneous in both observation and intrinsic physical properties such as stellar masses and redshifts.

\begin{figure*}[tb]
    \centering
    \begin{minipage}{0.49\textwidth}
        \centering
        \includegraphics[width=0.95\textwidth]{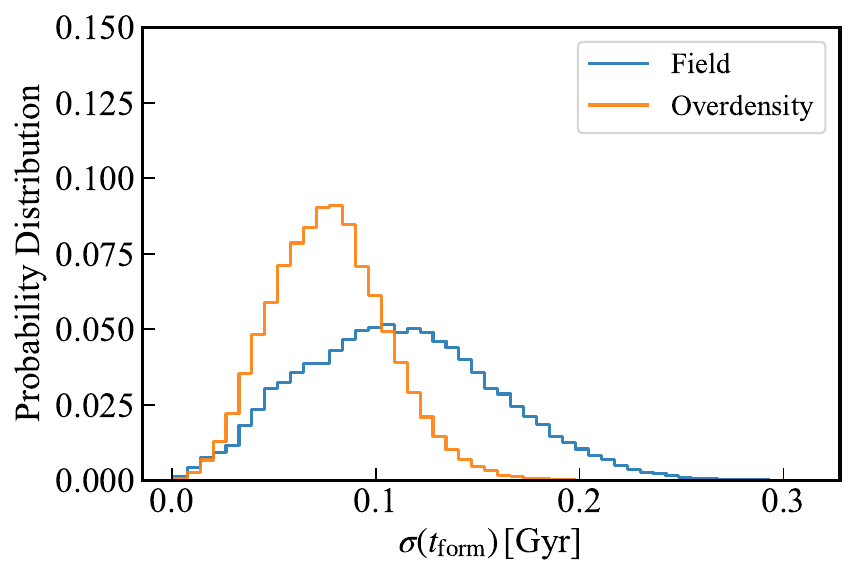}
    \end{minipage}
    \begin{minipage}{0.49\textwidth}
        \centering
        \includegraphics[width=0.95\textwidth]{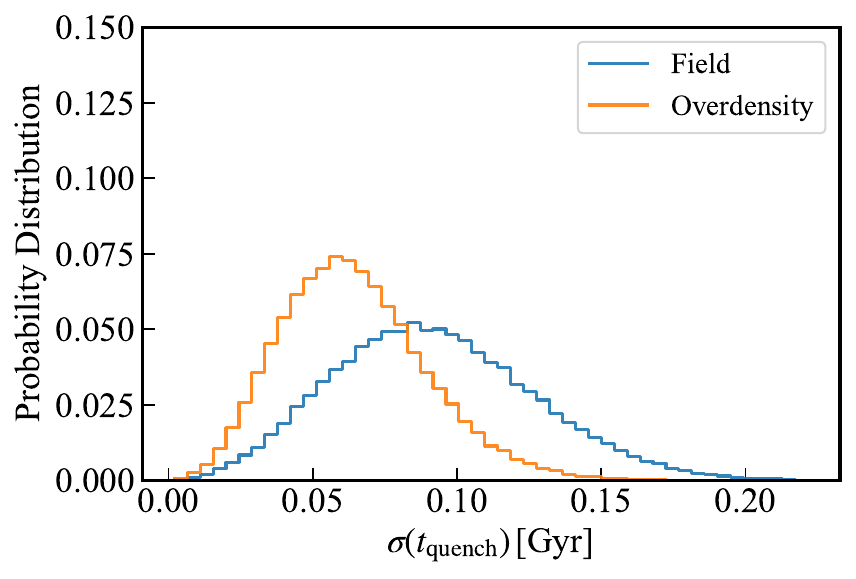}
    \end{minipage}
    \caption{Left Panel: Scatter of $t_\mathrm{form}$ based on 50000 times Monte Carlo calculation from the QGs within the overdensity (the orange histogram) and these in the field from \cite{DeepDive_2025} (the blue histogram). Right Panel: As in the left panel but for $t_\mathrm{quench}$. The figures indicate that QGs in the overdensities (cluster regions) form and quench at about the same time, whereas field QGs do not.}
    \label{fig:boot}
\end{figure*}

The left panel of Figure \ref{fig:boot} shows the distribution of the $t_\mathrm{form}$ scatter for 7 members in the overdensities compared to the randomly selected subsets of the field sample. The members show a smaller scatter ($\sigma_{t_\mathrm{form}}\sim 77\,\mathrm{Myr}$) compared to the randomly selected field samples ($\sigma_{t_\mathrm{form}} \sim 111\,\mathrm{Myr}$). 
Similarly, the right panel of the figure shows the $t_\mathrm{quench}$ distribution, reveling that the galaxies inside the overdensities have smaller scatter of quenching time (cluster: $\sigma_{t_\mathrm{quench}}\sim 62\,\mathrm{Myr}$, field: $\sigma_{t_\mathrm{quench}}\sim 90\,\mathrm{Myr}$). As one possible interpretation, these quantitative results support the scenario that QGs within the same overdensity formed and quenched at about the same time. In contrast, field QGs have more diverse formation and quenching epochs. This result is in line with theoretical work showing that galaxy environments are predictive of galaxy properties beyond number density \citep{Wu&Jespersen2023, Wu2024_environment}.

Previous studies have also reported the existence of a red sequence among massive QGs in proto-clusters in the early Universe \citep[e.g.,][]{Zirm2008,Ito2023,Tanaka2024}, which aligns with the similarity in formation times observed in this study. In particular, the member galaxies within the same overdensity exhibit this trend, indicating that large-scale environments, such as proto-clusters or large-scale structures, play a critical role in quenching massive galaxies at high redshift. For instance, \cite{Gonzalez2025} confirm the AGN component from \textit{Hyde} and surrounding sources, implying that major mergers can trigger AGN activity within this overdense environment. The other three galaxies around \textit{Jekyll \& Hyde} may experience higher rates of mergers and interactions compared to field QGs. The resulting feedback, particularly outflows from merger-induced AGNs \citep{Springel2005,Hopkins2006,Hopkins2008,Dubois2016}, may influence the rapid quenching of galaxies within this proto-cluster.

While this merger-driven scenario is only one of the possible interpretations of our results, it is consistent with the broader evolutionary picture of high-redshift overdensities.
In the high redshift Universe, many massive QGs have been confirmed to reside in overdensities \citep[e.g.,][]{Kakimoto2024, Carnall2024, Jespersen2025, deGraaff2025, Ito2025b}. \cite{McConachie2025} report that previously identified massive QGs also inhabit significant overdensities, suggesting that they assembled their stellar mass via merger-induced starbursts at even higher redshifts. Observationally, these intense starburst phases are identified as submillimeter galaxies (SMGs), which are widely considered to be the dust-obscured progenitors of massive QGs \citep[e.g.,][]{Toft_2014,Simpson2014,Gomez-Guijarro2018,Valentino2020,Dudzeviciute2020}. Supporting this evolutionary link, proto-clusters of SMGs have been discovered at $z > 4$ \citep{Miller2018,Mitsuhashi2021}, with some of them undergoing major mergers. Cosmological simulations further support this, predicting enhanced fractions of SMGs within proto-cluster cores \citep{ArayaAraya2024}. Together with these findings, our observations suggest that merger-induced starbursts and their subsequent quenching occur frequently, particularly in high-redshift overdensities.

In the local Universe, galaxies with older stellar populations tend to reside in high-density regions, implying an accelerated evolutionary path for cluster galaxies in the early Universe  \citep[e.g.,][]{Renzini_2006, Cooper2010, Webb2020}. Our results show that there are no significant differences in average stellar ages and quenching timescales between QGs in the overdensities and the field, which is consistent with the literature at high redshift \citep[e.g.,][]{Forrest2024}. Crucially, the fundamental difference lies not in the average age, but in the age distribution. While field massive QGs exhibit a wide range of stellar ages \citep[indicating stochastic, non-environmentally driven triggers; e.g.,][]{Schreiber_2018,Forrest_2020,carnall_2023,Setton2024,Wu2025,Nanayakkara2025,Baker2025_FLAMINGO}, the QGs within the overdense regions show a remarkably tight scatter in both formation and quenching times. This tight distribution suggests the presence of an environmental mechanism that drives the synchronized formation and subsequent quenching of these member galaxies, contrasting with the trends observed in the low-density field.

\subsection{Comparison with cosmological simulations} \label{subsec:TNG}
To examine whether current theoretical models can produce such extreme environments in the early Universe, we search for analogous overdense regions within a state-of-the-art cosmological simulation. We utilize the Illustris TNG300 simulation \citep{Pillepich2018,Nelson2019} and search for the overdensity at $z=3.71$ using the same method described in Section \ref{sec:target}. We define the R.A., Decl., and $z$ as `\texttt{pos\_x}', `\texttt{pos\_y}', and `\texttt{pos\_z}', respectively. The simulated analogs are defined as structures containing massive QGs within the comparable projected distance ($r_\mathrm{proj}<1.3\,\mathrm{pMpc}$) and line-of-site separation ($D_\mathrm{line}<8.5\,\mathrm{pMpc}$) to find the spatially equivalent systems. For overdensity calculations, we apply the redshift distribution of Figure \ref{fig:dens} (corresponding to 65.5\,pMpc) as a range of `\texttt{pos\_z}' (consistent with entire \texttt{pos\_z}). We note that if we change the coordinate information (such as `\texttt{pos\_y}' as $z$), the following results do not change. In addition, there are no clear overdensity of QGs at $z=4.01$ due to the low abundance of massive QGs \citep{Tanaka2024}.

\begin{figure}[tb]
    \plotone{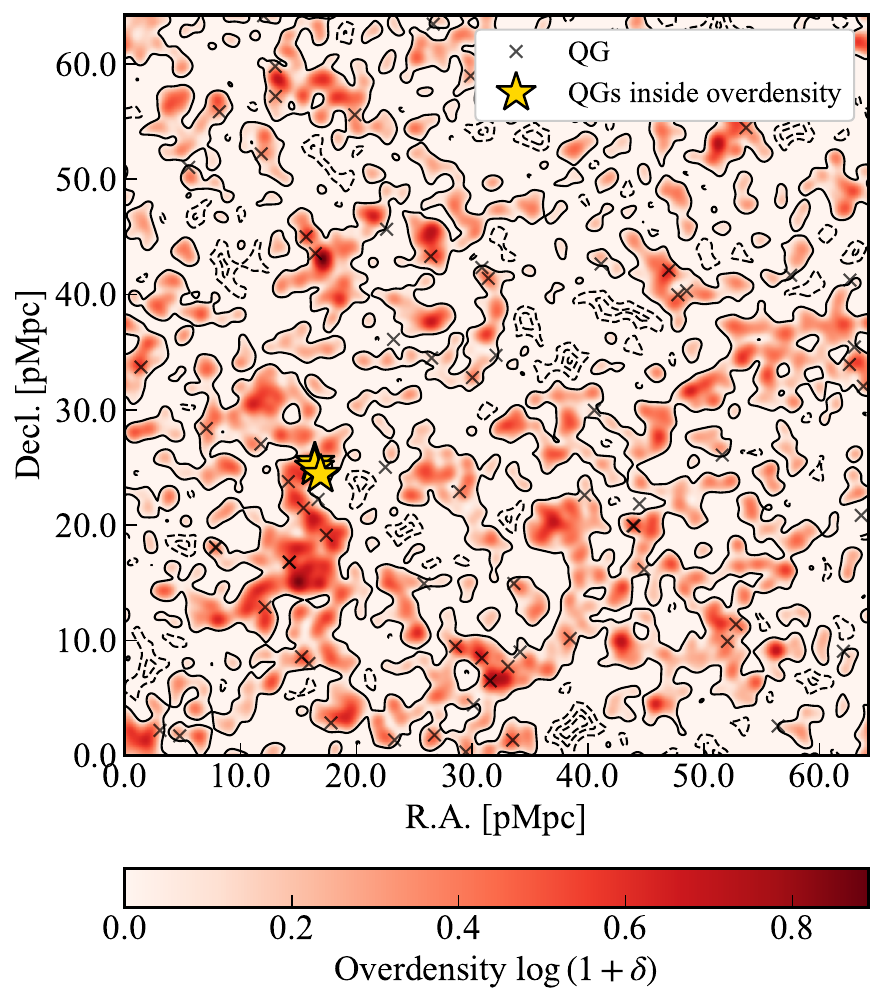}
    \caption{Overdensity significance of massive galaxies at $z=3.71$ (the red color map), smoothed over 500\,pkpc using Gaussian kernel density estimation in Illustris TNG300 as in the left panel of Figure \ref{fig:dens}. The yellow stars are QGs within the overdensity, and the black crosses indicate the positions of massive QGs within this map.
    \label{fig:TNGdens}}
\end{figure}

Figure \ref{fig:TNGdens} shows the overdensity of massive galaxies ($M_*>10^{10}\,M_\odot$) smoothed over 500\,pkpc using Gaussian kernel density inside the TNG300 (at $z=3.71$). In this figure, the overdensity value is computed using the same method as in Figure \ref{fig:dens}. The black crosses in Figure \ref{fig:TNGdens} show the distribution of massive QGs (defined as a galaxy with a star formation rate less than 10\% of the SFMS), and there is an overdensity with three massive QGs (the yellow stars, $\texttt{SubhaloID}=110, 111, 184$, hereafter QG1, QG2, and QG3, respectively) centered at $\mathrm{R.A.} \sim 15\,\mathrm{pMpc}$ and $\mathrm{Decl.} \sim 25\,\mathrm{pMpc}$ with 3.8 times more galaxies than the average ($\delta \sim 2.8$). This figure indicates that while the overdensity observed around \textit{Jekyll \& Hyde} is reproduced, only a single overdensity region with several QGs is detected across the simulation box. This is also the highest redshift overdensity which consists of more than two massive QGs. We note that while the peak overdensity is higher than that in Figure \ref{fig:dens}, the overall density contrast in the simulation is stronger because observational structures are smoothed out by photometric redshift uncertainties.

Intriguingly, the QGs evolve to the brightest cluster galaxy (BCG). We trace the descendants of the QGs in the overdensity to $z=0$ using the merger trees in the simulation. The three QGs merge at $z=2.10$ (QG1 and QG2) and $z=1.30$ (QG2 and QG3), and the descendant at $z=0$ is identified as the central galaxy of its host halo ($\texttt{SubhaloID}=33174$). Furthermore, the total mass of the host halo reaches $M_{200} \sim 10^{14.8}\ M_\odot$, strictly satisfying the definition of a BCG \citep{Pillepich2018, Montenegro-Taborda2023}. This suggests that the QGs around \textit{Jekyll \& Hyde} could merge into a single galaxy in the local Universe, which suggests that the overdensity is proto-BCG. 

\begin{figure*}[tb]
    \plotone{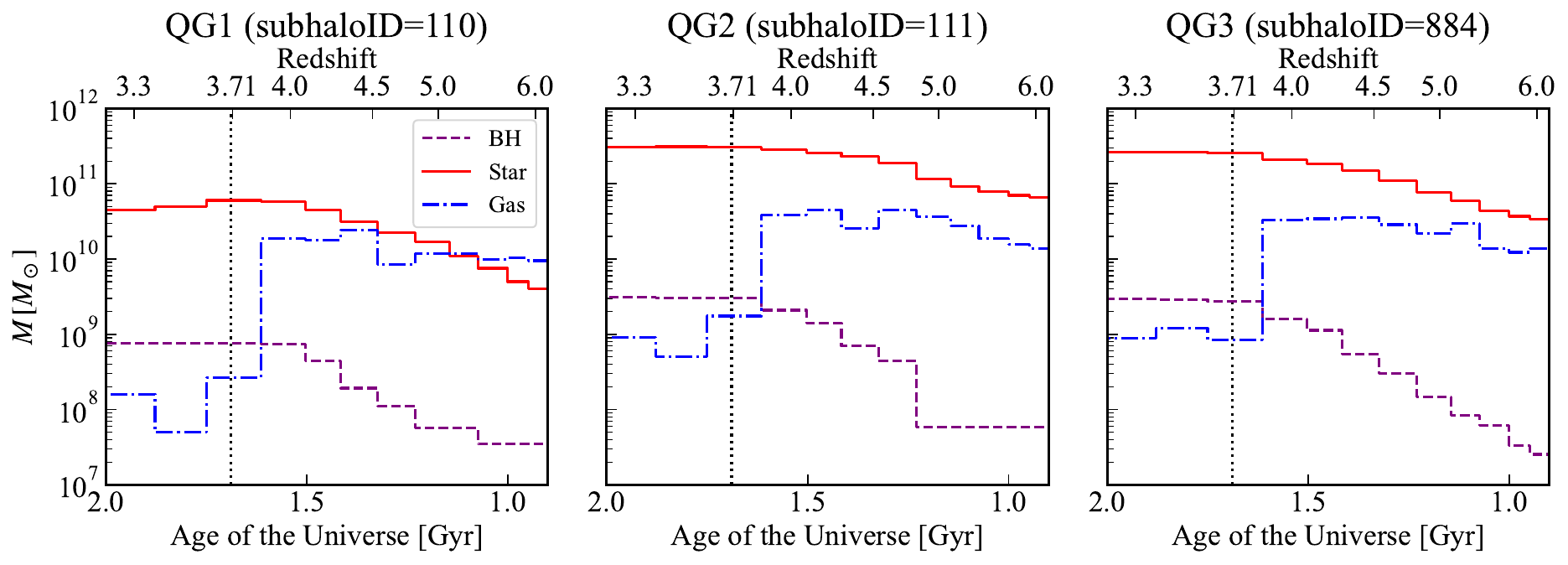}
    \plotone{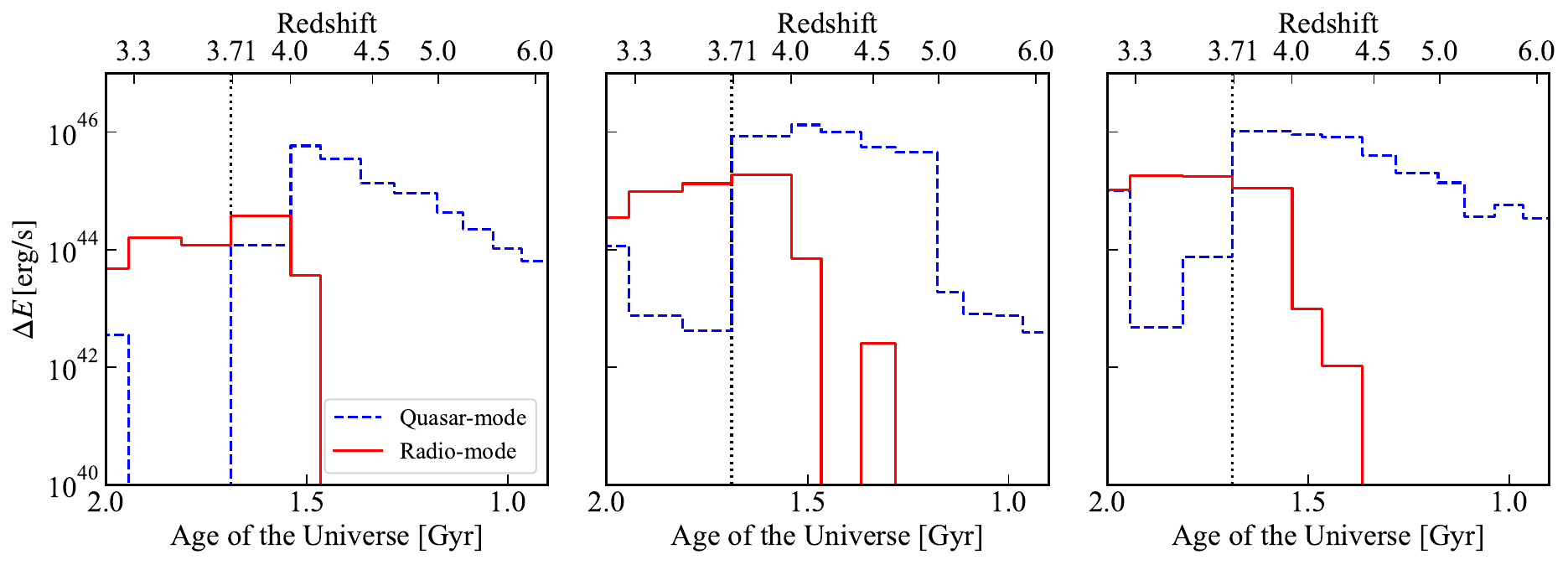}
    \caption{Top Panels: Mass assembly history of massive QGs within the overdensity in Illustris TNG300 ($z=3.71$, $\texttt{SubhaloID}=110, 111, 184$). The purple dashed line, the red solid line, and the blue dashdotted line indicate stellar mass, black hole mass, and gas mass, respectively. Bottom Panels: History of energy injected by quasar-mode and radio-mode AGN feedback (the red solid line and the blue dashed lines, respectively). The values represent the time-averaged energy injection between consecutive snapshots from $z=6.0$ to $z=3.0$.
    \label{fig:TNGAGN}}
\end{figure*}

Figure \ref{fig:TNGAGN} shows the assembly history of stellar mass, gas mass, and black hole mass of the QGs within the overdensity at each snapshot. All QGs stopped their star formation activity in the same snapshot (at $z=3.71$), which is also consistent with the observation, and their gas mass rapidly decreased in that snapshot. This suggests that the quenching is initiated by the depletion of surrounding star-forming gas, which is the main fuel for star formation. This is due to the existence of a super-massive black hole (SMBH). All QGs have a SMBH with $M_\mathrm{bh} \sim 10^9\,M_\odot$, and they cause the radio and quasar mode feedback for the immediate surroundings of galaxies \citep[see bottom of Figure \ref{fig:TNGAGN};][]{Terrazas2020}. Especially, the energy from radio (kinetic) mode feedback, which ejects the gas around the galaxies, is rapidly increased just before the snapshot (the red line) and drives galaxy quenching \citep{Hartley2023,Kurinchi-Vendhan2024,Lagos2025}. Thus, although limited to a single system, the successful reproduction of a QG overdensity at the same redshift suggests that AGN outflows may play a role in their synchronous quenching.
Another interesting point is that two of the three QGs (QG1 and QG2) experienced major mergers at $z= 4.43$ and $z= 4.66$, respectively, which accelerates the evolution of $M_\mathrm{bh}$. This is in line with the evolution of AGN, one of the physical mechanisms for quenching in galaxies, being highly influenced by galaxy mergers, as has been suggested theoretically \citep{Jespersen2022_mergers, Chuang2024_mergers}.

While the formation of massive QGs via merger-induced AGN activity and/or starbursts is observed in several cosmological simulations, the universal role of the environment in galaxy quenching remains debated. Several studies emphasize that QGs and their starburst progenitors (e.g., bright SMGs) experience frequent mergers and are more abundant in overdense regions \citep{Xie2024,ArayaAraya2025}. Conversely, other research indicates that this type of evolution does not strictly depend on host halo mass, implying that QGs can form in various environments \citep[e.g.,][]{ArayaAraya2025model, DeLucia2025, Baker2025_FLAMINGO}. Although our results suggest that large-scale environments are essential for driving the synchronized evolution of QGs in overdense regions, the similarities in average quenching timescales between member QGs and field QGs indicate that the fundamental quenching mechanism, such as AGN feedback, operates universally across both dense and field environments at high redshift. Another caveat is that different simulations implement the AGN feedback in different ways; the quenching epoch and timescale will be dependent on the detailed recipe of AGN feedback. Our results here are only for Illustris TNG300.

\subsection{Molecular gas and galaxy quenching} \label{subsec:gas}
Molecular gas mass surrounding the QGs around \textit{Jekyll \& Hyde} is already constrained by the previous research \citep{Suzuki_2022}. Band-3 observation did not detect [C\,{\footnotesize I}] line from \textit{Jekyll} (DD-134), DD-129, and ZF-COS-18842, which suggests the molecular gas mass fraction is lower than 20\% \citep[also consistent with the literature; e.g.,][]{Gobat2018, Scholtz2026}. The observed low gas contents of QGs indicate that the depletion of gas material around QGs may be important for suppressing their star formation activity.

The scarcity of molecular gas in observations is the equivalent of the lack of star-forming gas in simulation results, which might further support the scenario of merger-induced AGN activity and subsequent quenching. On the other hand, however, studies comparing theoretical models with the SFHs of high-redshift QGs argue that merger events are not strictly necessary to reproduce massive galaxies in the early Universe; rather, large amounts of accreting gas from the surrounding large-scale overdensity are crucial \citep{Kurinchi-Vendhan2024, Jespersen2025, deGraaff2025, Traina2025}. In either case, our results highlight the importance of the large-scale environment for triggering the galaxy quenching and suggest that this environment is also critical for the evolution of central SMBHs in the early Universe. Member galaxies within these overdensities may simultaneously experience quenching driven by AGN feedback, leading to the formation of proto-clusters populated by multiple QGs in the early Universe. Furthermore, these overdensities may become the progenitors of BCGs in the local Universe through the merging of their member QGs, which is also observed in high redshift pairs of QGs \citep[e.g.,][]{Ito2025b}.

\section{Conclusion} \label{sec:conclusion}
In this study, we spectroscopically confirm two galaxy overdense regions around \textit{Jekyll \& Hyde} at $z=3.71$ and SXDS-27434 at $z=4.01$. The follow-up observation with JWST/NIRSpec in \textit{DeepDive} and JWST archival sample confirms three massive ($M_*>10^{10}\,M_\odot$) QGs around \textit{Jekyll} and two massive QGs around SXDS-27434. They show similar star formation histories. Comparison of the formation and quenching redshifts between the QGs in the overdensities and field QGs suggests that the QGs within the same overdensity are formed and quenched about the same time. This trend indicates that large-scale environments such as galaxy proto-clusters and/or large-scale structures may have an important role in quenching of massive galaxies at this redshift.

These galaxies may experience more mergers and interactions than field QGs, which may invoke outflows from merger-induced AGN and drive galaxy quenching in this overdensity. In fact, the Illustris TNG300 simulation reproduces one overdensity of massive QGs at $z=3.71$ (same redshift as one of our overdensities). Our analysis of this overdensity suggests that merger-induced AGN deplete the molecular gas surrounding of galaxies, which causes synchronous quenching of the QGs. These merger-induced activities and/or large amounts of gas accretion facilitated by their massive host halos may drive the simultaneous formation of several massive galaxies and central SMBHs within overdensities at high redshifts.
There are now many QGs confirmed by JWST in the early Universe, but it is quite difficult to confirm the surrounding environment using only photometric information. Future spectroscopic follow-up observations around these QGs are thus crucial to confirm their formation mechanisms and their surrounding environments.

\begin{acknowledgments}
We thank the anonymous referee for constructive comments. TK acknowledges support from JSPS grant 25KJ1331. This work was partially supported by Overseas Travel Fund for Students (2025) of Astronomical Science Program and The Graduate University for Advanced Studies, SOKENDAI. FV, KI, and PZ acknowledge support from the Independent Research Fund Denmark (DFF) under grant 3120-00043B. JAD is supported by a Dunlap Fellowship, funded through an endowment established by the David Dunlap family and the University of Toronto. WMB gratefully acknowledges support from DARK via the DARK fellowship. This work was supported by a research grant (VIL54489) from VILLUM FONDEN. DC was supported by the Agencia Estatal de Investigación (AEI, Spain)  under projects PID2024-156100NB-C21 and CNS2024-154550, funded by MCIN/AEI/10.13039/501100011033. MF and MH acknowledge funding from the Swiss National Science Foundation (SNF) via a PRIMA Grant PR00P2193577 `From cosmic dawn to high noon: the role of black holes for young galaxies'. MO acknowledges support from JSPS KAKENHI Grant Number 25K07361. J.R.W. acknowledges that support for this work was provided by The Brinson Foundation through a Brinson Prize Fellowship grant. This work is based [in part] on observations made with the NASA/ESA/CSA James Webb Space Telescope. The data were obtained from the Mikulski Archive for Space Telescopes at the Space Telescope Science Institute, which is operated by the Association of Universities for Research in Astronomy, Inc., under NASA contract NAS 5-03127 for JWST. These observations are associated with program \#2565 and \#3567. Some of the data products presented herein were retrieved from the Dawn JWST Archive (DJA). DJA is an initiative of the Cosmic Dawn Center (DAWN), which is funded by the Danish National Research Foundation under grant DNRF140.
\end{acknowledgments}





%
\facilities{JWST (NIRSpec)}

\software{astropy \citep{2013A&A...558A..33A,2018AJ....156..123A,2022ApJ...935..167A},  
          \texttt{emcee} \citep{2013PASP..125..306F},
          Matplotlib \citep{Hunter:2007},
          \texttt{MIZUKI} \citep{Tanaka_2015},
          numpy \citep{harris2020array},
          \texttt{prospector} \citep{Johnson_2021},
          \texttt{pysersic} \citep{Pasha2023},
          Python-fsps \citep{Conroy_2009,Conroy_2010}, 
          Scipy \citep{Virtanen2020},
          Source Extractor \citep{1996A&AS..117..393B}
          }






\bibliography{sample7}{}
\bibliographystyle{aasjournalv7}



\end{CJK*}
\end{document}